\documentclass[structabstract]{aa7}  
\usepackage{graphicx}
\usepackage{txfonts}
\usepackage{natbib}

\bibpunct{(}{)}{;}{a}{}{,} 

\begin{document}
 \title{\textit{Chandra}\thanks{The \textit{Chandra} data described in this paper
have been obtained in the open time project with Sequence Number 200767 and
ObsID 13613 (PI: T.~Preibisch).}
X-ray observation of the \ion{H}{ii} region Gum~31 in the Carina nebula 
complex\thanks{Tables 1 and 3 are only available in electronic form
at the CDS via anonymous ftp to cdsarc.u-strasbg.fr (130.79.128.5)
or via http://cdsweb.u-strasbg.fr/cgi-bin/qcat?J/A+A/}
}

   \author{T.~Preibisch\inst{1} \and M.~Mehlhorn\inst{1}  
\and L.~Townsley\inst{2}
\and P.~Broos\inst{2} 
\and T.~Ratzka\inst{1,3}
          }

   \institute{Universit\"ats-Sternwarte M\"unchen, Ludwig-Maximilians-Universit\"at,
          Scheinerstr.~1, 81679 M\"unchen, Germany \email{preibisch@usm.uni-muenchen.de} 
\and
Department of Astronomy \& Astrophysics,
             Pennsylvania State University, University Park PA 16802, USA
\and
Institute for Physics / IGAM, Karl-Franzens-Universit\"at, 
Universit\"atsplatz 5/II, 8010 Graz, Austria}

\titlerunning{\textit{Chandra} X-ray observation of Gum~31}
\authorrunning{Preibisch, Mehlhorn, Townsley, Broos, Ratzka}

\date{Received 26 November 2013; accepted 10 March 2014}

 
  \abstract
   {Gum~31 is a prominent, but still rather poorly studied
 \ion{H}{ii} region around the stellar cluster NGC~3324 at the
northwestern periphery of the Carina nebula complex. 
   }
   {Our aim was to reveal and characterize the young stellar population
 in Gum~31. 
An X-ray survey is the only efficient
way to identify young stars in this region with extremely high
galactic field-star contamination that can avoid the strong biases
of infrared excess selected samples of disk-bearing young stars.
   }
   {We used the \textit{Chandra} observatory
   to perform a deep (70 ksec) X-ray observation of the Gum~31 region
 and detected 679 X-ray point sources. 
This extends and complements the X-ray survey of the central Carina nebula
regions performed in the \textit{Chandra Carina Complex Project} (CCCP).
Using deep near-infrared images
from our recent VISTA survey of the Carina nebula complex,
our comprehensive \textit{Spitzer} point-source catalog, and
optical archive data, we identify counterparts for 75\% of 
these X-ray sources.
   }
   {The spatial distribution of the X-ray selected young stars
shows two major concentrations,
the central cluster NGC~3324 and a partly
embedded cluster in the southern rim of the \ion{H}{ii} region.
However, these two prominent clusters contain only 
about 30\% of the X-ray selected population, whereas the majority
($\sim 70\%$) of X-ray sources constitute 
a rather homogeneously distributed population of young stars.
Our color-magnitude diagram analysis suggests ages of $\sim 1-2$~Myr
for the two clusters, whereas the distributed population shows a wider
age range up to $\sim 10$~Myr.
We also identify previously unknown companions to two of the three
O-type members of NGC~3324 and detect diffuse X-ray emission in two 
parts of the region.
   }
 {An extrapolation based on the observed X-ray luminosity function 
suggests that the observed region contains about 4000 young stars in total
(down to $0.1\,M_\odot$).
This shows that the Gum~31 area contains 
a substantial fraction of the total stellar
population in the CNC. 
The distributed population of young stars in the
Gum~31 region is probably a part or extension 
of the widely distributed population of $\sim 1 - 10$~Myr old stars,
that was identified in the CCCP area. This implies that the global
stellar configuration 
of the Carina nebula complex is a very extended
stellar association, in which the (optically prominent) clusters
contain only a minority of the stellar population.
}


   \maketitle
%

\section{Introduction}

The Carina nebula complex (CNC)
is one of the largest, most massive, and most active
star-forming complexes in our Galaxy. Located at a moderate
and well known distance of 2.3~kpc \citep{Smith06}, 
the nebulosity extends
over about 100~pc, corresponding to several degrees on the sky.
The clouds contain a total gas and dust mass of about $10^6\,M_\odot$
\citep{Preibisch12} and harbor
presumably about 100\,000 young stars 
\citep{CCCP-Clusters,Povich11,HAWKI-survey}.
The star formation rate of
$\sim 0.01 - 0.02\,M_\odot/{\rm yr}$ \citep[see][]{Povich11,Gaczkowski13} constitutes 
as much as about 1\% of the total galactic star formation.
The large population of massive stars 
\citep[$\ge 70$ O-type and WR stars][]{Smith06}, including
the object $\eta$~Car (which is the most luminous known star in our
Galaxy)
creates very high levels of ionizing radiation and stellar wind power,
which profoundly influence the surrounding clouds.
This stellar feedback has already dispersed a large fraction of 
the original dense molecular clouds, out of which the stars formed, but
the compression of the remaining clouds by the ionization fronts
and expanding wind bubbles is also currently triggering
 the formation of new generations of stars in the complex \citep{Smith10b,Gaczkowski13}.

A general review of the CNC is provided in the book chapter by
\citet{SB08}.
During the last five years, several new and sensitive surveys
of different parts of the CNC have been performed at wavelengths
from the X-ray to submm regime.
One major milestone was the deep X-ray imaging survey of the
{\it Chandra} Carina Complex Project \citep[CCCP; see][for an overview]{CCCP-intro},
which mapped an area of about 1.4 square-degrees with a mosaic of 22  individual
ACIS-I pointings, using a total observing time of
1.34 Mega\-seconds (15.5 days).
With a detection limit corresponding to
X-ray luminosities of about $10^{30}\,\rm erg/s$, these X-ray data 
can detect the coronal X-ray emission of young ($\la 10^7$ yrs) stars down to 
$\sim 0.5\,M_\odot$.
The \textit{Chandra} images revealed
14\,368 individual X-ray sources, and a
sophisticated classification scheme
showed that 10\,714 of these
are most likely young stars in the Carina nebula \citep{CCCP-classification}.
The analysis of the spatial distribution of the X-ray detected young stars showed
that half of the young stellar population resides in one of about 30 clusters or stellar
groups, while the other half constitutes a widely dispersed population 
\citep{CCCP-Clusters} that is spread throughout the entire observed area.
The combination of these CCCP X-ray data with a deep near-infrared
survey \citep{HAWKI-survey} obtained with HAWK-I at the ESO VLT
provided information about the properties of the stellar populations
in the central parts of the complex \citep{CCCP-HAWKI},
including the individual stellar
clusters Tr~16 and Tr~15 \citep{CCCP-Tr16,CCCP-Tr15}. 

Because of the wide spatial extent of the CNC, most recent studies (and all 
those mentioned above) have focused on the central 
$\la 1.5$~square-degree area.
At the northwestern periphery of the CNC, about
$80'$ (or $\approx 50$~pc) from $\eta$~Car,
the \ion{H}{ii} region Gum~31 \citep[first described by][]{Gum55}
constitutes the most
prominent object in this area of the sky (see Fig.~1).
Gum~31 is a roughly circular nebula with a diameter 
of about $12'$ ($\approx 8$~pc),  and
is excited by the young stellar cluster NGC~3324
 (containing three known O-type stars).
Despite its interesting morphology and the publicity of
\textit{Hubble Space Telescope}\footnote{{\tt hubblesite.org/newscenter/archive/releases/2008/34/} }
 and ground-based optical images\footnote{e.g., 
ESO Photo Release \,\,{\tt \footnotesize www.eso.org/public/news/eso1207/}},
the Gum~31 nebula (sometimes designated as the Gabriela Mistral nebula)
remained quite poorly studied until very recently,
probably because the closeness to the extremely
eye-catching  Carina nebula has always overshaded Gum~31.
In the cluster NGC~3324, 
stellar spectral types are only known for
the three optically brightest stars, HD~92\,206 A, B, and C,
which have been classified as O6.5V (HD~92\,206 A and HD~92\,206 B)
and O9.5V  (HD~92\,206 C) \citep{Mathys88,Walborn82}.
\citet{Carraro01}
identified 25 candidate members by means of optical photometry
and suggested
that the cluster age is presumably a few Myr.
In addition to the optically visible cluster NGC~3324 
in the \ion{H}{ii} region, a few dozen of embedded young star candidates
have been found in the clouds surrounding the \ion{H}{ii} region,
either in infrared images
\citep{Cappa08} or traced by their protostellar jets 
\citep{Smith10a,Ohlendorf12}.
However, these objects known
until recently must represent only the tip of the iceberg:
given the presence of three O-type stars ($M \ge 18\,M_\odot$), the 
extrapolation of the canonical stellar IMF \citep{Kroupa02} would
suggest a total population of $\approx 1500$ low-mass ($0.1\,M_\odot \le M \le 2\,M_\odot$) stars.
It is thus quite obvious that the
vast majority of the stellar population in the Gum~31 region
is still completely unknown.

While \citet{Walborn82} already concluded that the distance
modulus values of the brightest stars in NGC~3324 are consistent with 
those of the clusters Tr~16 / Col 228 in the central
parts of the Carina nebula, the apparent segregation of the
Gum~31 nebula from the bright H$\alpha$ emission of the Carina nebula
in optical images left the physical relation unclear.
New information on this question came from our recent \textit{Herschel} 
far-infrared observations \citep{Preibisch12,Roccatagliata13} of the CNC.
The field-of-view of this survey was
wide enough (more than 5 square-degrees) to 
include the area around Gum~31. The \textit{Herschel} maps
showed that the dense dusty shell surrounding the Gum~31 \ion{H}{ii} region 
is connected by numerous filamentary cloud structures
to the dense molecular clouds in the inner parts of the
Carina nebula, suggesting Gum~31 to be part of the CNC. 
The shell around Gum~31 and the surrounding clouds 
host several dozens of deeply embedded
protostars and pre-stellar cores, which are detected
as point-like sources in our \textit{Herschel} data \citep{Gaczkowski13}.
In the recent study of \citet{Ohlendorf13}, these
\textit{Herschel} data were combined with \textit{Spitzer} archive data and 
the Wide-field Infrared Survey Explorer (WISE) point-source catalog
and lead to the identification of some 600 young stellar object candidates
in a one square-degree  area centered on Gum~31
by means of their infrared excesses.
A clear concentration of partly embedded young stellar objects is located
in a dense cloud at the southwestern edge of the Gum~31 bubble.
This very young cluster, which we will designate as G286.38--0.26 in following text,
is located at a position where the Gum~31 shell seems to
interact with the expanding super-bubbles driven by the numerous
massive stars in the central parts of the CNC.
This may represent an interesting example of triggered star formation
in a cloud that formed and/or is compressed by colliding large-scale shocks.

The current, infrared-excess selected 
sample of protostars and disk-bearing young stars in the Gum~31 region,
resulting from our analysis of the \textit{Herschel} and
\textit{Spitzer} data \citep{Ohlendorf13},
contains objects down to stellar masses of about $1\,M_\odot$.
However, since the stars in this region 
are presumably already several Myr old, 
this infrared-excess selected sample must be highly incomplete, because
it is well known that the typical lifetime of 
circumstellar disks around young stars are just a few Myr
\citep[e.g.,][]{Fedele10}. 
The young stars which have already dispersed their disks will
no longer exhibit infrared excesses and can thus not be identified
by  infrared excess selection.
This represents a major obstacle in the identification of the 
young stellar population in the Gum~31 region, which is a fundamental
prerequisite for an determination of its star formation history.
Furthermore, because
Gum~31 lies very close to the galactic plane 
($b \approx -0.2^\circ$) a great deal of confusion results 
from galactic field star contamination.
All optical and infrared images of this region are thus completely
dominated by unrelated field stars, and
it is therefore impossible to identify and distinguish a population
of several Myr old low-mass stars from
unrelated field stars with optical or infrared photometry alone.

An X-ray survey, however, can  solve this problem:
the strongly enhanced
X-ray emission of young ($\la 10^7$~yrs) stars 
\citep[see, e.g.,][]{Feigelson07,Preibisch_coup_orig} 
provides an extremely useful
discriminant between
young pre-main sequence stars and the much older field stars.
The median X-ray luminosity of a few Myr old
solar-mass stars is nearly 1000 times higher than
for solar-mass field stars \citep[see][]{PF05},
and makes these young stars relatively easily detectable targets
for current observatories, even at the relative large distance of 2.3~kpc.
This was the motivation to perform the deep {\em Chandra} X-ray
observation of the Gum~31 region presented in this paper.
In Section 2 we describe the X-ray
observation and data analysis and discuss basic properties of the
X-ray sources. We then describe in Sect.~3 the deep near-infrared data
from our recent VISTA survey of the CNC and our \textit{Spitzer} point-source
catalog that are then used in Sect.~4 
to identify and characterize the counterparts of the  X-ray sources.
In Sect.~5 we discuss the X-ray and infrared/optical
properties of particularly interesting objects.
In Sect.~6 we use near- and mid-infrared color-color and color-magnitude 
diagrams to infer the extinctions, infrared excesses, masses, and ages
of the X-ray detected objects.
Section 7 contains a discussion of the global properties of the X-ray selected
population in Gum~31. In Sect.~8 we present and briefly discuss the
diffuse X-ray emission that is detected in our {\em Chandra} image.
Finally, section 9 contains a brief summary and the conclusions from this
study.

   \begin{figure*}
 \parbox{9cm}{\includegraphics[width=9cm]{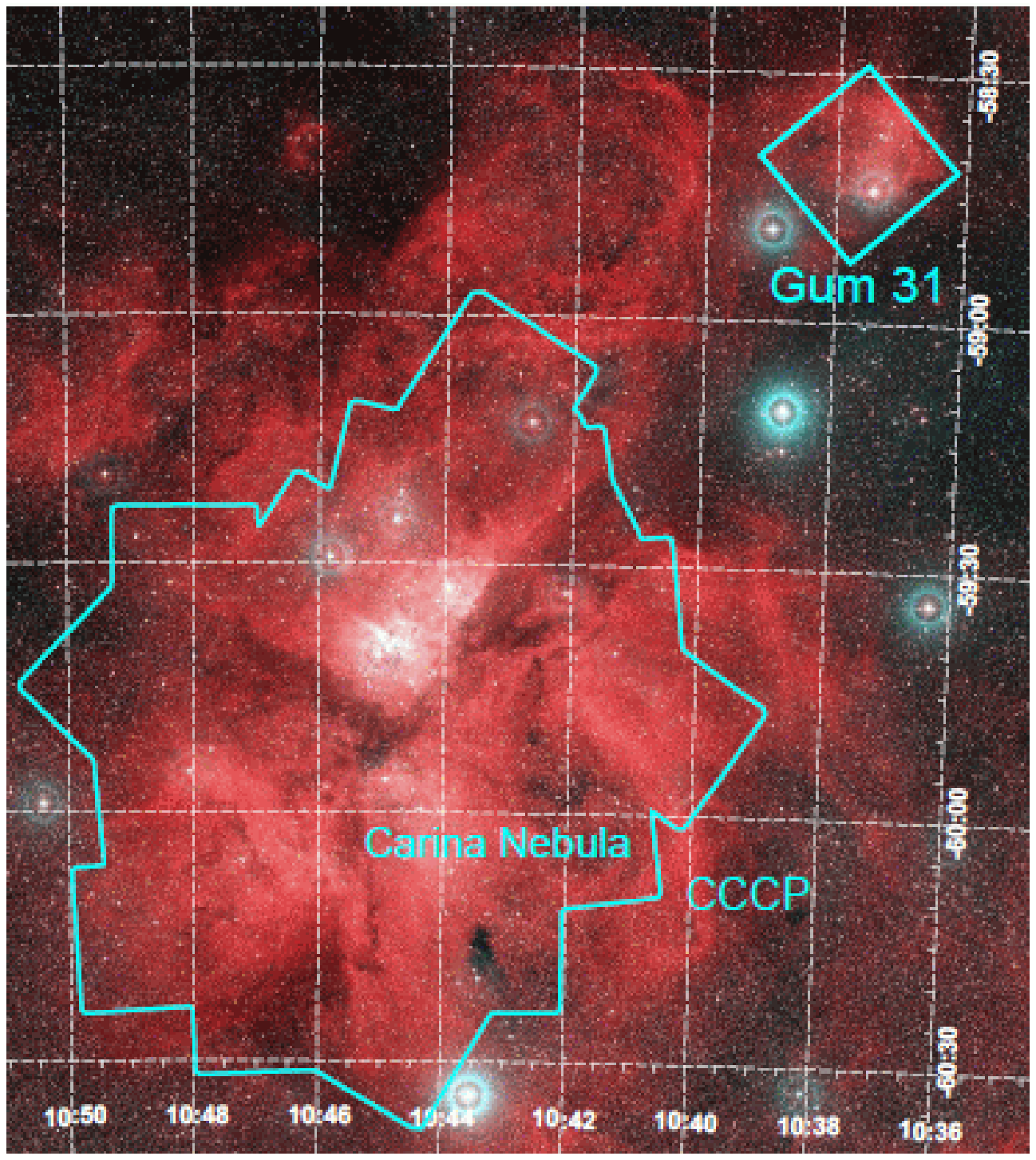}}
 \parbox{9.2cm}{\includegraphics[width=9.2cm]{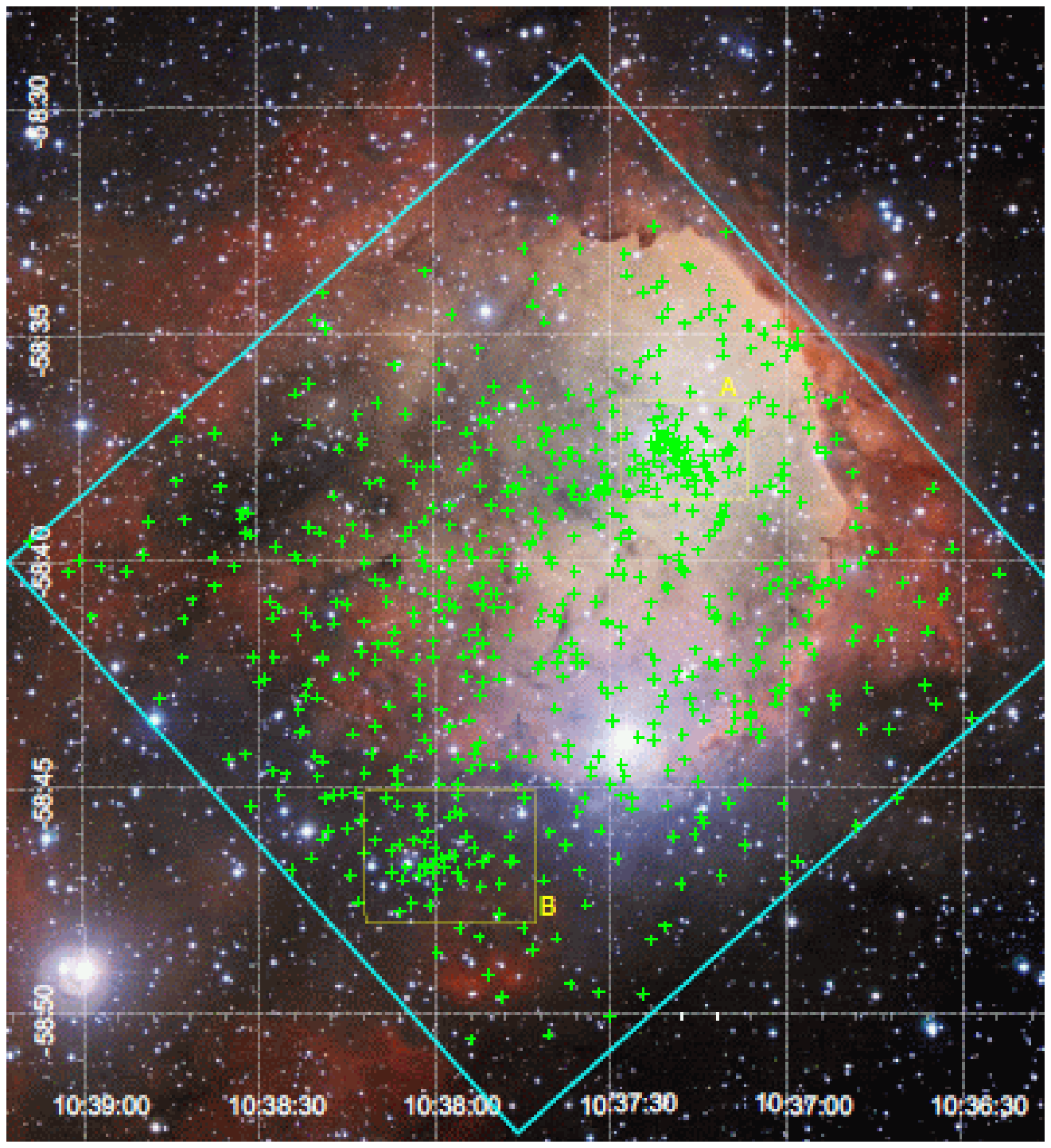}}
   \caption{Left: Optical image of the Carina nebula complex
(from: {\tt www.eso.org/public/images/eso0905b/}; image credit: ESO/Digitized Sky Survey 2, Davide De Martin).
The region observed in the context of the \textit{Chandra} Carina Complex Project (CCCP) and 
the \textit{Chandra} pointing of the Gum~31 region are marked by the cyan outlines.\newline
Right: Optical image of the Gum~31 region obtained with the Wide Field Imager on the MPG/ESO 
2.2~m 
telescope at La Silla Observatory (from: {\tt www.eso.org/public/images/eso1207a/}; image credit: ESO) 
with outline of the
\textit{Chandra} pointing in cyan. The individual X-ray point sources are marked by green crosses.
The yellow boxes mark the 
two regions shown in Fig.~\ref{n3324-g286.fig} (box {\sf A} corresponds to the
region of the cluster NGC~3324, while box {\sf B} corresponds to the region of the cluster G286.38--0.26)
  }
              \label{outlines.fig}%
    \end{figure*}

\section{\textit{Chandra} X-ray observation and data analysis}

We have used the \textit{Chandra} observatory \citep{Weisskopf02} 
 to perform a deep pointing of the Gum~31 region with the
Imaging Array of the \textit{Chandra}
Advanced CCD Imaging Spectrometer \citep[ACIS-I; see][]{Garmire03}.
The observation was performed as an open time project 
with Sequence Number 200767 and
ObsID 13613 (PI: T.~Preibisch)
during \textit{Chandra} Observing Cycle~13 in 
October 2012 (start date: 2012-10-08T20:20:42, 
end date: 2012-10-09T16:43:01).
ACIS-I provides a field of view of $17' \times 17'$ on the sky\footnote{Although two CCDs
of the spectroscopic array ACIS-S were also operational during our pointing,
we did not include these data in our analysis since the point-spread function is 
seriously degraded at the corresponding large offaxis angles. Inspection of the 
two ACIS-S array regions shows a high level of almost homogeneously background, but
no obvious point sources.}
(what corresponds to  $11.3 \times 11.3$~pc at the distance of
2.3~kpc),  and has a pixel size of $0.492''$.
The point spread function of the X-ray telescope
has a FWHM of $0.5''$ on-axis, but
increases towards the edge of the detector.
Since the \textit{Chandra} aspect reconstruction errors 
are usually very small ($\la 0.1''$),
the positions of bright X-ray sources can usually be determined
with subarcsecond precision, and 
sources with separations as close as $\la 1''$ can be resolved.

The aimpoint of the observation was set to be
  $\alpha ({\rm J2000}) = 10^{\rm h}\,37^{\rm m}\,36.6^{\rm s}$,
$\delta ({\rm J2000}) = -58\degr\,41'\,18''$.
This position is close to the center of the \ion{H}{ii} region, and
allows both the stellar cluster NGC~3324 and
the cluster G286.38--0.26 to be in the inner parts of the field-of-view, where the
point-spread function is still very good.
The pointing roll angle
(i.e.,~the orientation of the detector with respect to the celestial 
north direction) was $138.35\degr$.
The ACIS field-of-view is just wide enough to cover the
full spatial extent of the optically bright Gum~31 \ion{H}{ii} region
and some parts of the surrounding dust shell 
(see Fig.~1).

The observation was performed in the standard
``Timed Event, Very Faint'' mode with $5 \times 5$ pixel event islands.
The total net exposure time of the observation
was 68\,909~s (19.14 hours).

At the distance of 2.3~kpc, the expected ACIS point source sensitivity limit
for a 5-count detection on-axis in a 70~ks observation is
$L_{\rm X,min} \sim  10^{29.8}$~erg~s$^{-1}$,
assuming an extinction of 
$A_V \leq 2.5$~mag ($N_{\rm H} \leq 5 \times 10^{21}\,{\rm cm}^{-2}$)
as typical for the stars in the \ion{H}{II} region,
and a thermal plasma
with $kT = 1$~keV \citep[which is a typical value for young stars; see, e.g.,][]{Preibisch_coup_orig}.
Using the  empirical relation between X-ray luminosity and stellar mass
and the temporal evolution of X-ray luminosity
from the 
sample of young stars in the Orion nebula Cluster
that was very well studied in the \textit{Chandra} Orion 
Ultradeep Project \citep{Preibisch_coup_orig,PF05},
we can expect to detect most ($> 80\%$) of the stars with 
$M \ge 1\,M_\odot$
and about half of the $0.1\, M_\odot < M < 1\,M_\odot$ stars in a
$\leq 5$~Myr old population.
These estimates are applicable
to the lightly-obscured stars in NGC~3324. 
For the more deeply embedded sources in the denser clouds,
assuming $A_V = 7.5$~mag ($N_{\rm H} = 1.5 \times 10^{22}\,{\rm cm}^{−2}$)
and a thermal plasma with $kT = 3$~keV (typical for embedded young stellar objects),
the sensitivity limit increases only slightly to 
$L_{\rm X,min}\approx 10^{30.0}$~erg~s$^{-1}$.
These limits are very similar to those of the CCCP observations of the central parts 
of the Carina nebula.

\subsection{X-ray point-source detection and analysis with \textit{ACIS Extract}}

All details of the employed data analysis techniques are discussed 
in \citet{Broos10}, hereafter referred to as B10; 
we just provide a brief summary of this procedure here. 
\textit{Chandra}-ACIS event data were calibrated and cleaned as described 
in Section 3 of B10. Those procedures seek to improve the accuracy of 
event properties (individual event positions, alignment of the 
\textit{Chandra} coordinate system to an astrometric reference, 
energy calibration) and seek to discard events that are likely caused 
by various instrumental background components. 

Candidate point sources were then identified in the pointing using two methods,
the standard \textit{Chandra} detection tool, {\tt wavdetect} \citep{Freeman02}
and a Lucy-Richardson image reconstruction \citep{Lucy74}, that were performed
on images in different energy bands and pixels sizes.
These candidate sources were then extracted using the 
{\em ACIS Extract} (hereafter AE) software 
package\footnote{ The {\em ACIS Extract} software package and User's Guide are available 
at {\tt http://www.astro.psu.edu/xray/acis/acis\_analysis.html}. } \citep{AE2012}.

For each source, AE calculated the probabilities that the counts extracted in each of 
three energy bands arose solely from the local background.
When all three probabilities were greater than 0.01 or when less than three X-ray counts 
were extracted, we judged the candidate source to be not significant, and removed it from the list.
The positions of surviving source candidates were updated with AE estimates, and the 
reduced list of candidates was re-extracted.
This cycle of extraction, pruning, and position estimation was repeated until no candidates 
were found to be insignificant.

Our final X-ray catalog contains 679 individual point sources.
The number of extracted counts ranges
from 3 for the faintest sources, up to 920 for the strongest source;
the median value is 11 counts.
AE screened all sources for an instrumental non-linearity known as 
{\em photon pile-up}\footnote{\tt http://cxc.harvard.edu/ciao/why/pileup\_intro.html};
no sources were found to be at risk of pile-up.

The final list of X-ray sources with their properties
is reported in Table~1 \onltab{1}{}
(available in the electronic edition). Sources are sorted
by increasing right ascension and identified by their sequence number (Col.~1)
or their IAU designation (Col.~2). 
Following the rules for the
designation of sources found with the \textit{Chandra} X-ray Observatory,
we have registered the acronym 
\textit{CXOPMTB} as the prefix for the IAU designation (Col.~2).

\subsection{X-ray source variability}

   \begin{figure*}
   \centering
 \parbox{14.5cm}{\includegraphics[width=7.0cm]{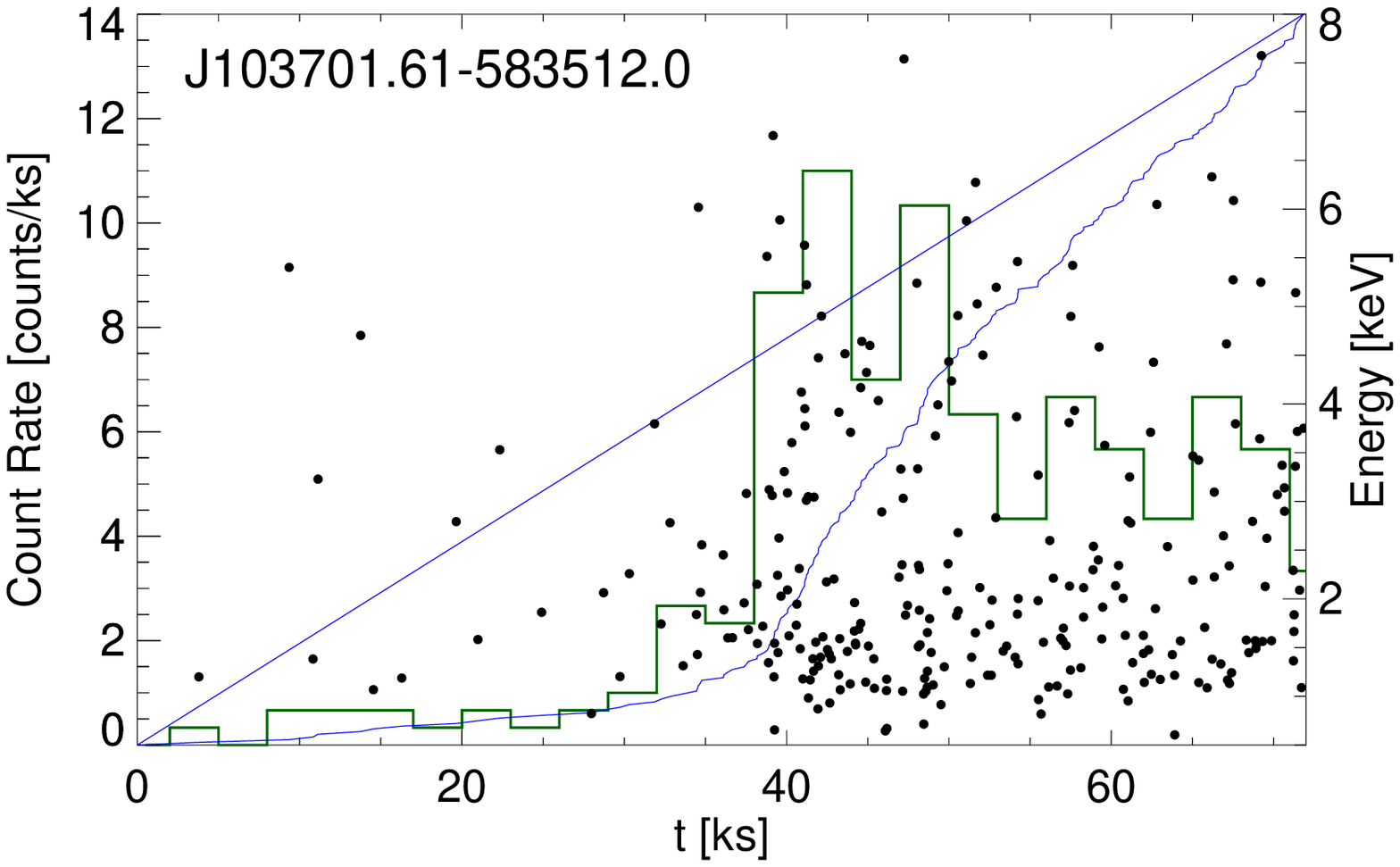} \hspace{4mm}
   \includegraphics[width=7.0cm]{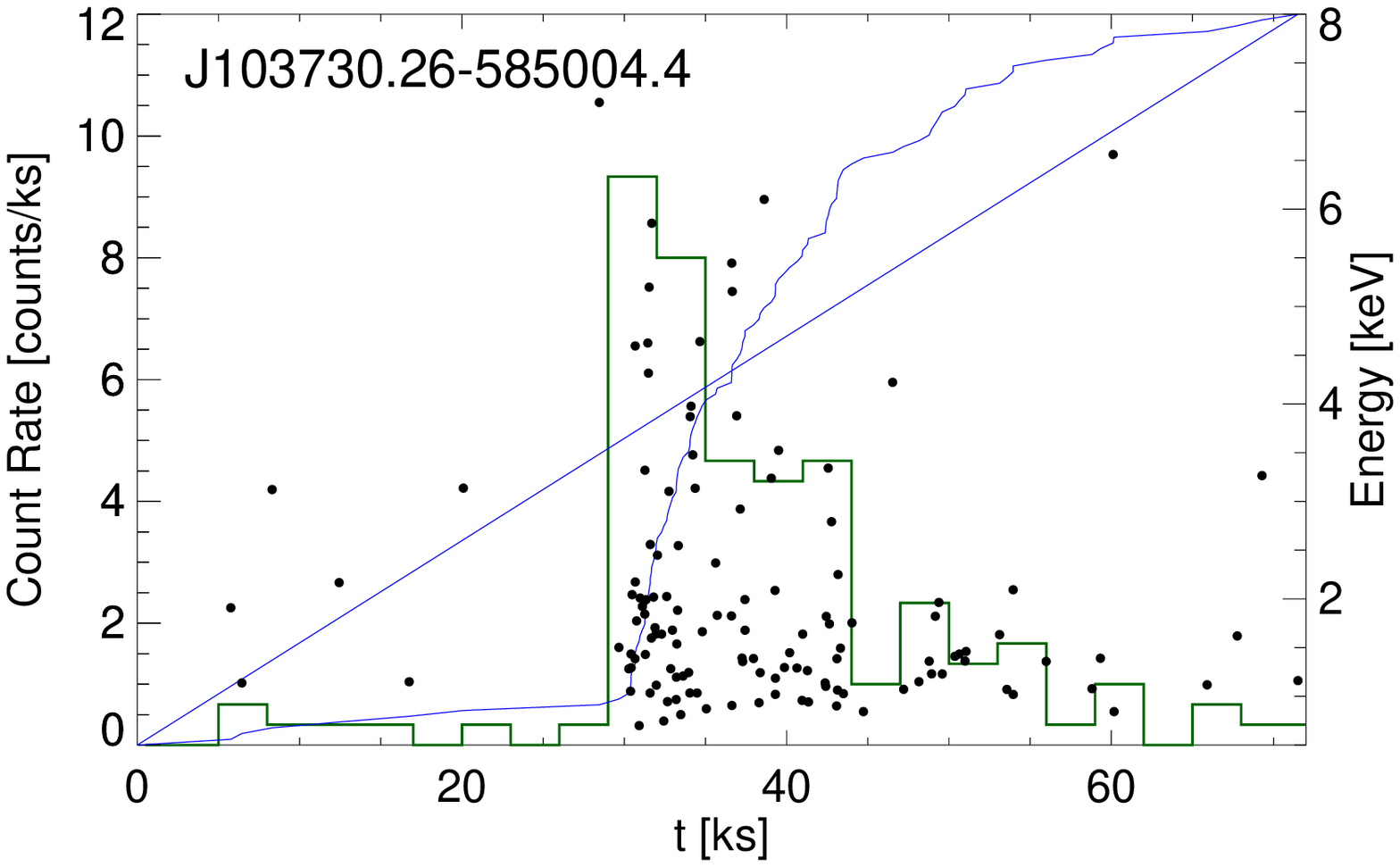}} \\
 \parbox{14.5cm}{ \includegraphics[width=7.0cm]{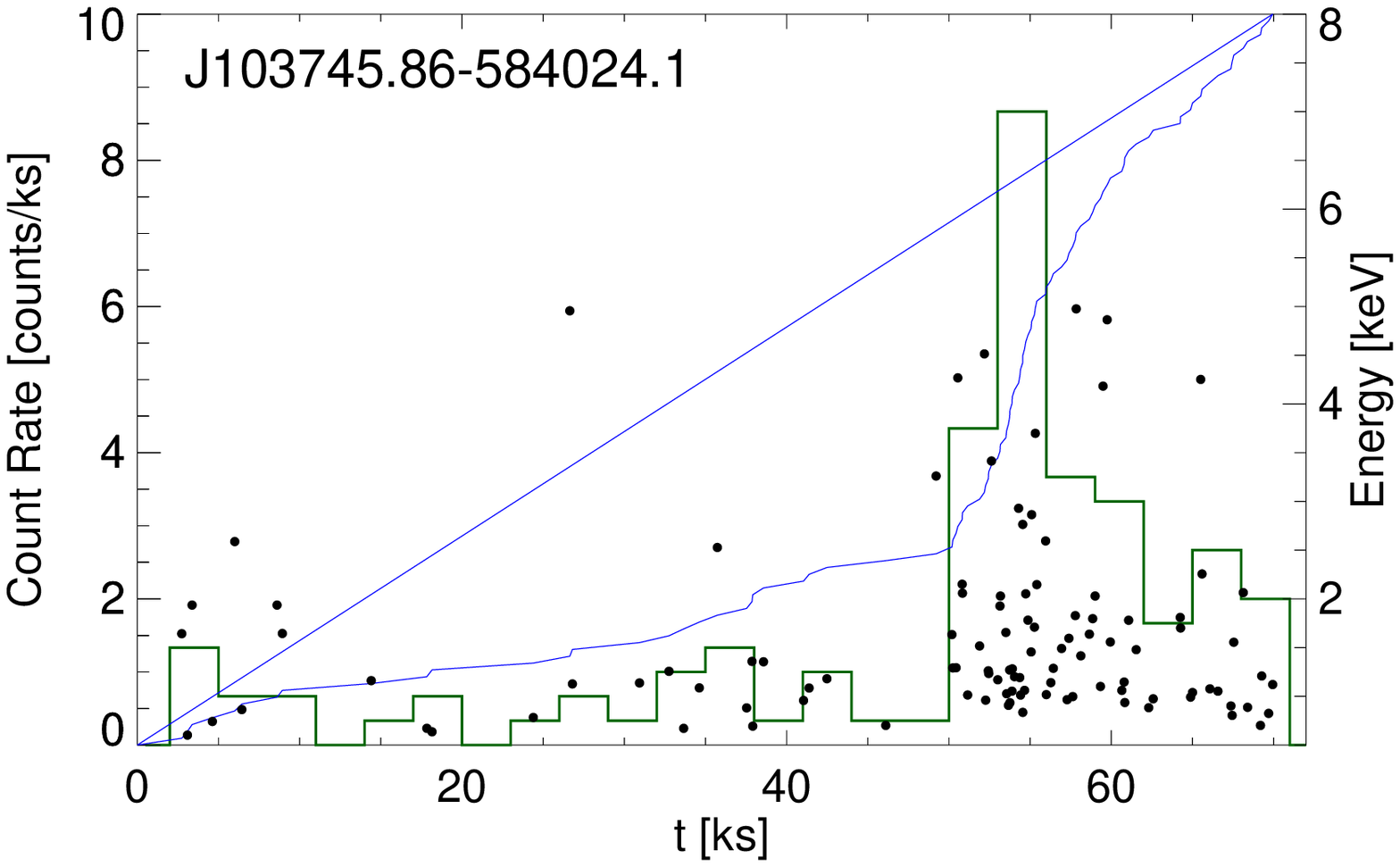}\hspace{4mm}
   \includegraphics[width=7.0cm]{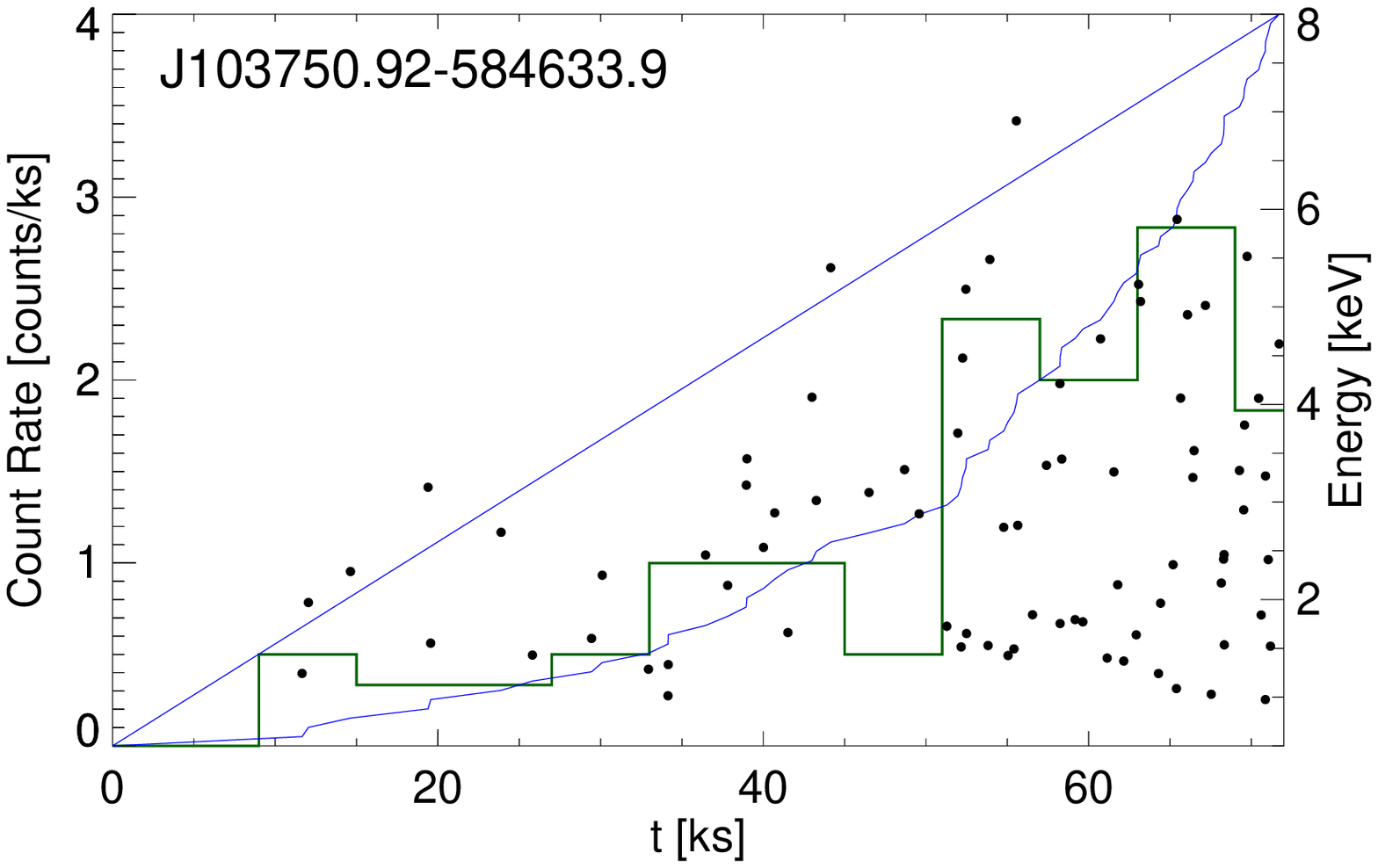}}\\
 \parbox{14.5cm}{  \includegraphics[width=7.0cm]{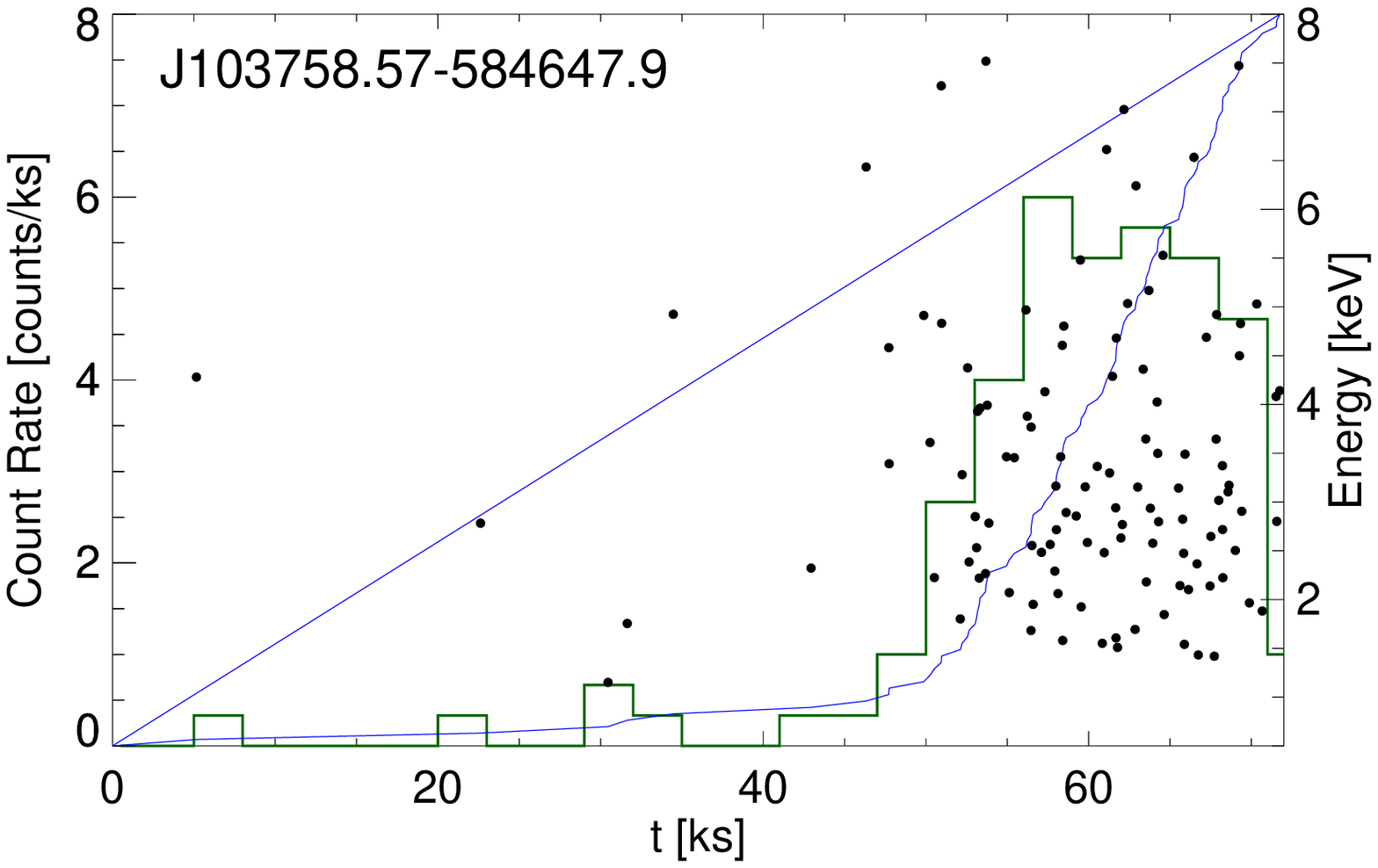} \hspace{4mm}
   \includegraphics[width=7.0cm]{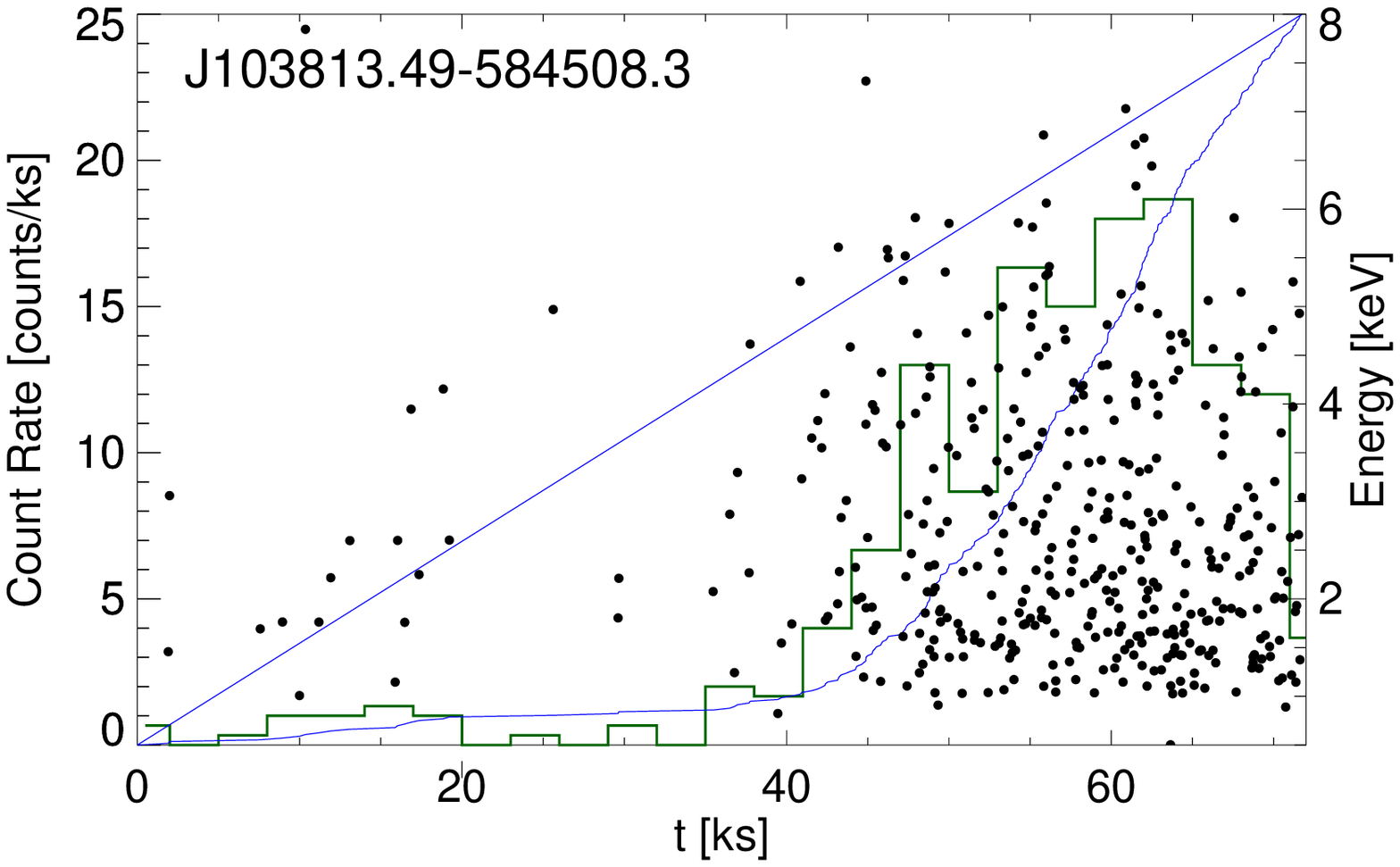} }
   \caption{Lightcurves for six significantly variable sources. The solid dots
    show the arrival time (measured from the start of the observation)
    and the energy of each of the detected source photons. The histograms show
    the corresponding binned lightcurves. The solid lines show the
    cumulative distribution function of the photon arrival times
    compared to the expectation for a perfectly constant
  source (the diagonal line).
   }
              \label{fig:lightcurves}%
    \end{figure*}

The AE procedure investigates the time variability of each X-ray source
by comparing the arrival times of the individual
source photons in each extraction region to a model assuming
temporal uniform count rates.
The statistical significance for variability 
is then computed with a 1-sided Kolmogorov-Smirnov
statistic (Col.~15 of Table~1). In our sample,
 42 sources show significant X-ray variability (probability of being constant
$P_{\rm const} <0.005$) and additional 48 sources
 are classified as possibly variable 
($0.005 <  P_{\rm const} < 0.05$).

The light curves of the variable X-ray sources show
a variety of temporal behavior; six of the most interesting lightcurves
are shown in Fig.~\ref{fig:lightcurves}. Three of these sources show
flare-like variability, i.e.,~a fast increase in the count rate followed
by a slow exponential decay, as typical for
solar-like magnetic reconnection flares \citep[see, e.g.,][]{Wolk05}.
The other variable sources show
more slowly increasing or decreasing count rates, as also often found
for young stellar objects
\citep[see, e.g.,][]{Stassun06}.

\subsection{X-ray spectral fits of bright sources} 
\label{ssec:spec}

   \begin{figure*}
   \centering
\parbox{14.5cm}{\includegraphics[width=6.5cm,bb = 31 26 540 415, clip]{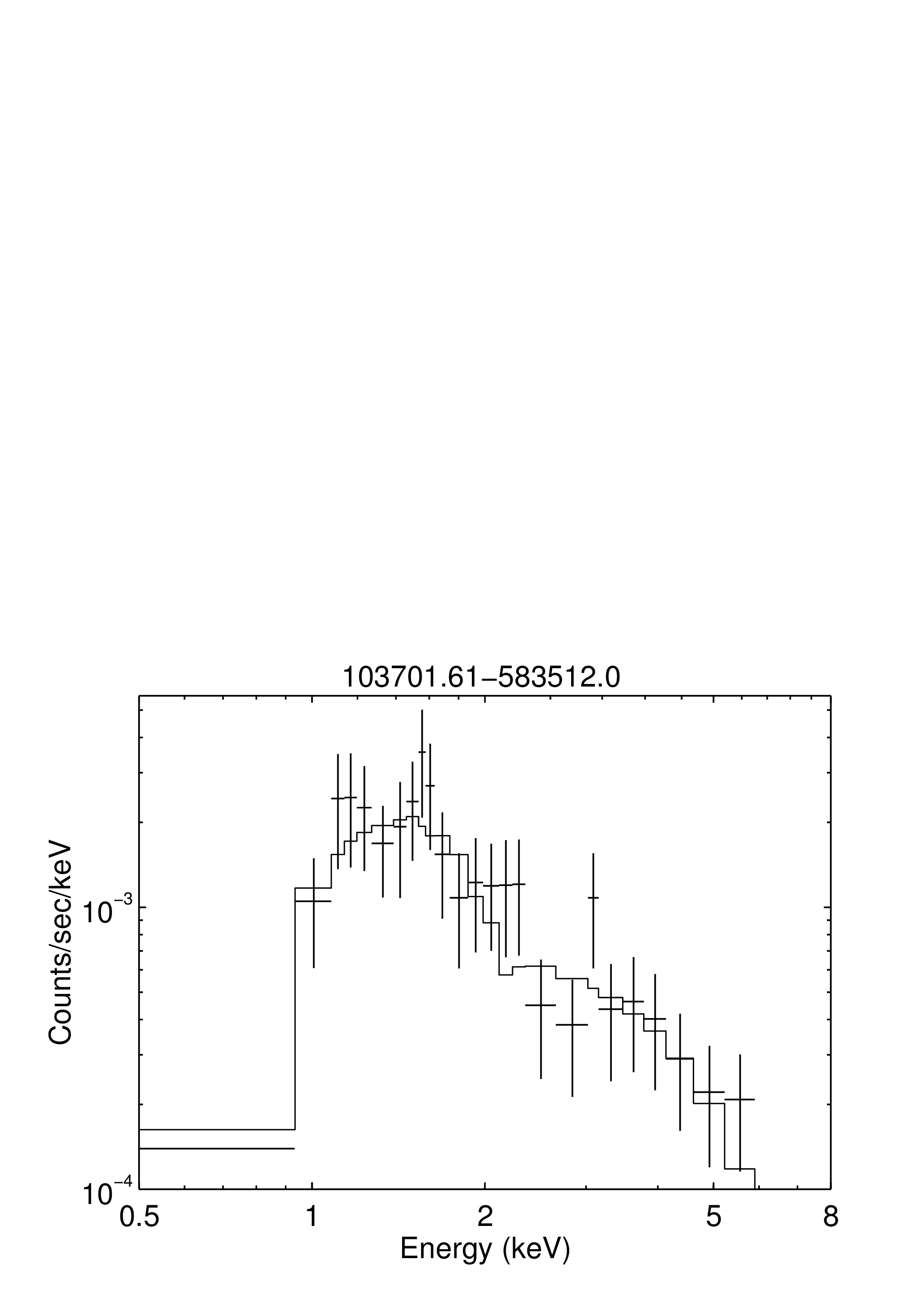} \hspace{4mm}
\includegraphics[width=6.5cm,bb = 31 26 540 415, clip]{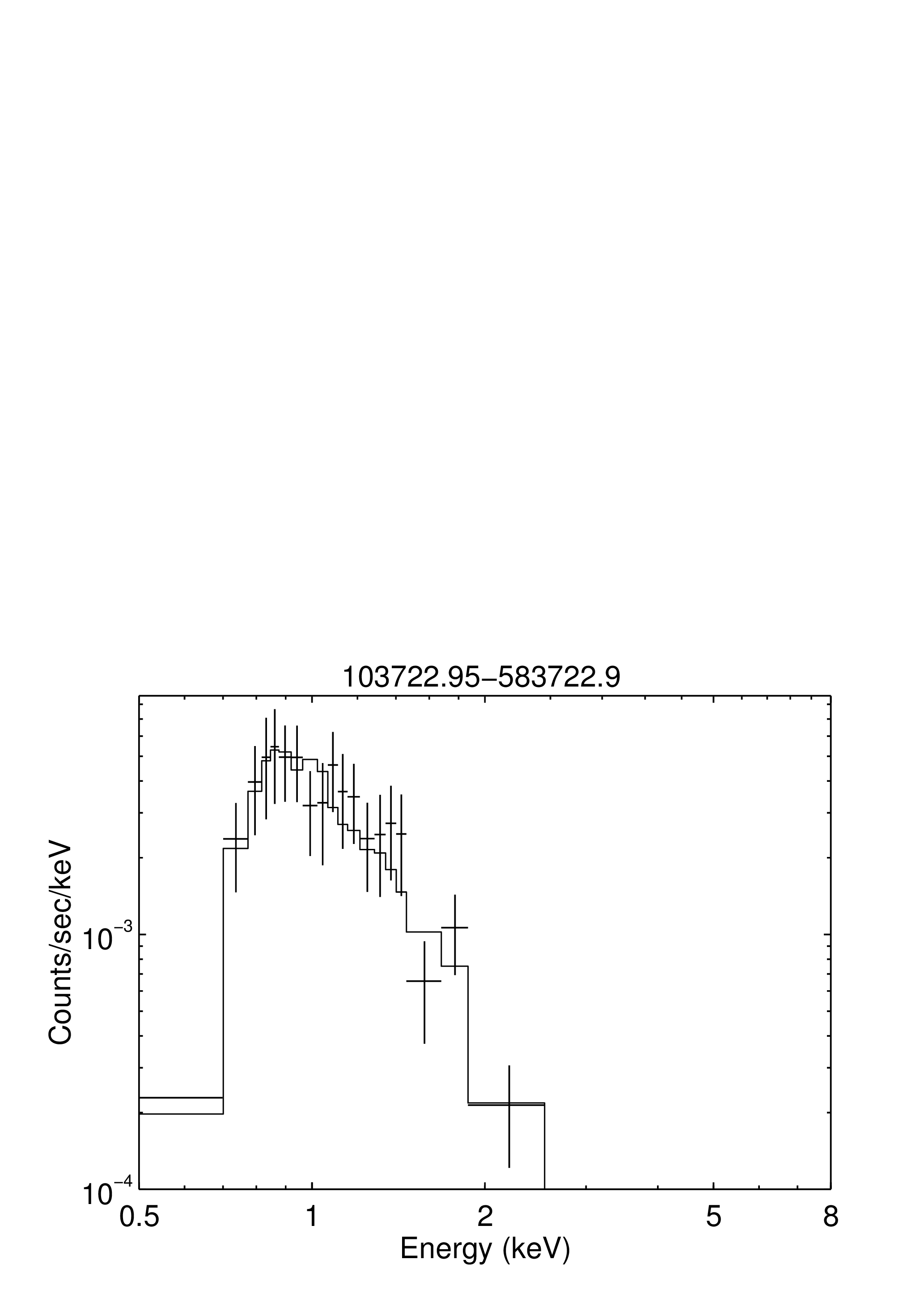}}\\
\parbox{14.5cm}{\includegraphics[width=6.5cm,bb = 31 26 540 415, clip]{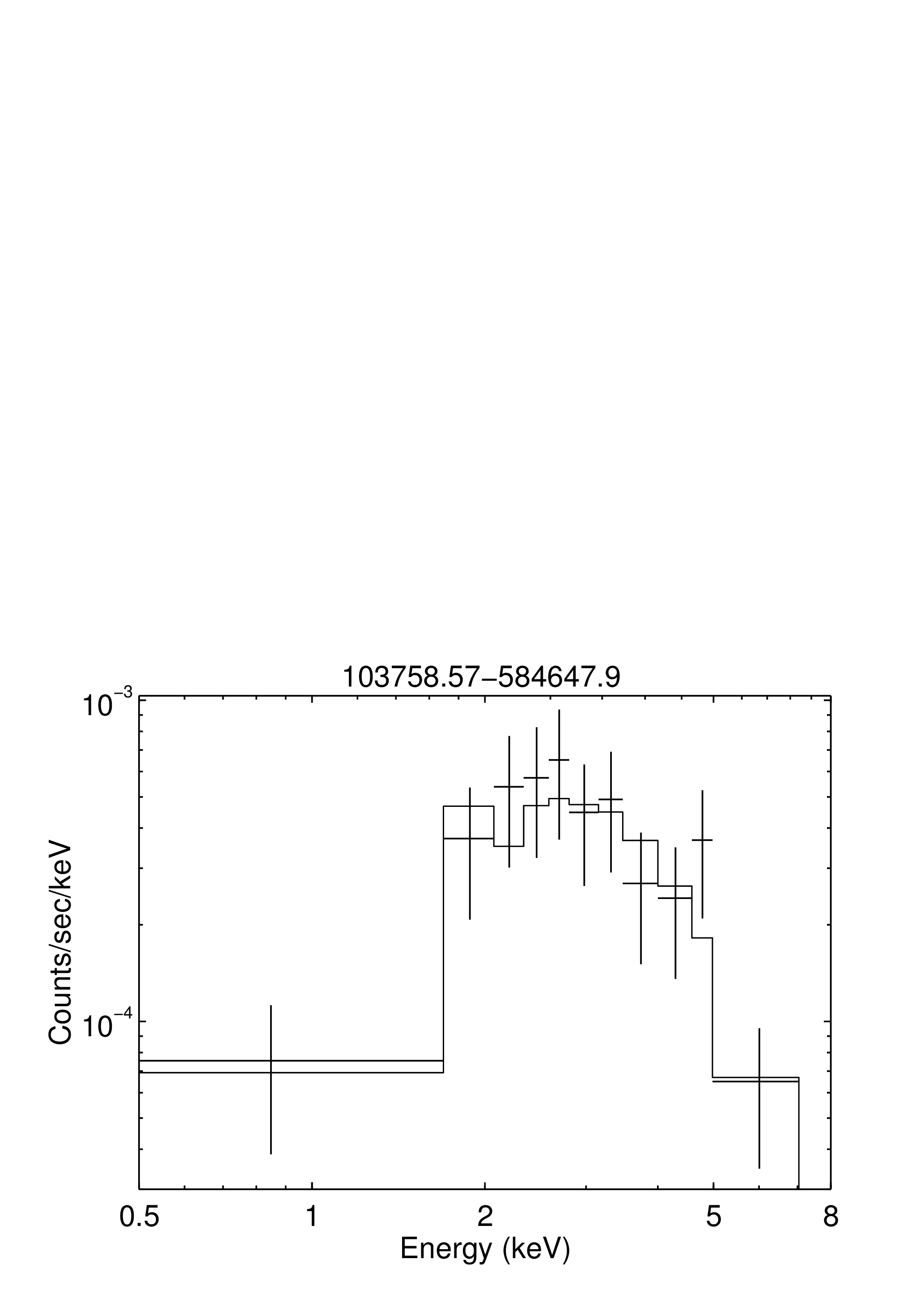} \hspace{4mm}
\includegraphics[width=6.5cm,bb = 31 26 540 415, clip]{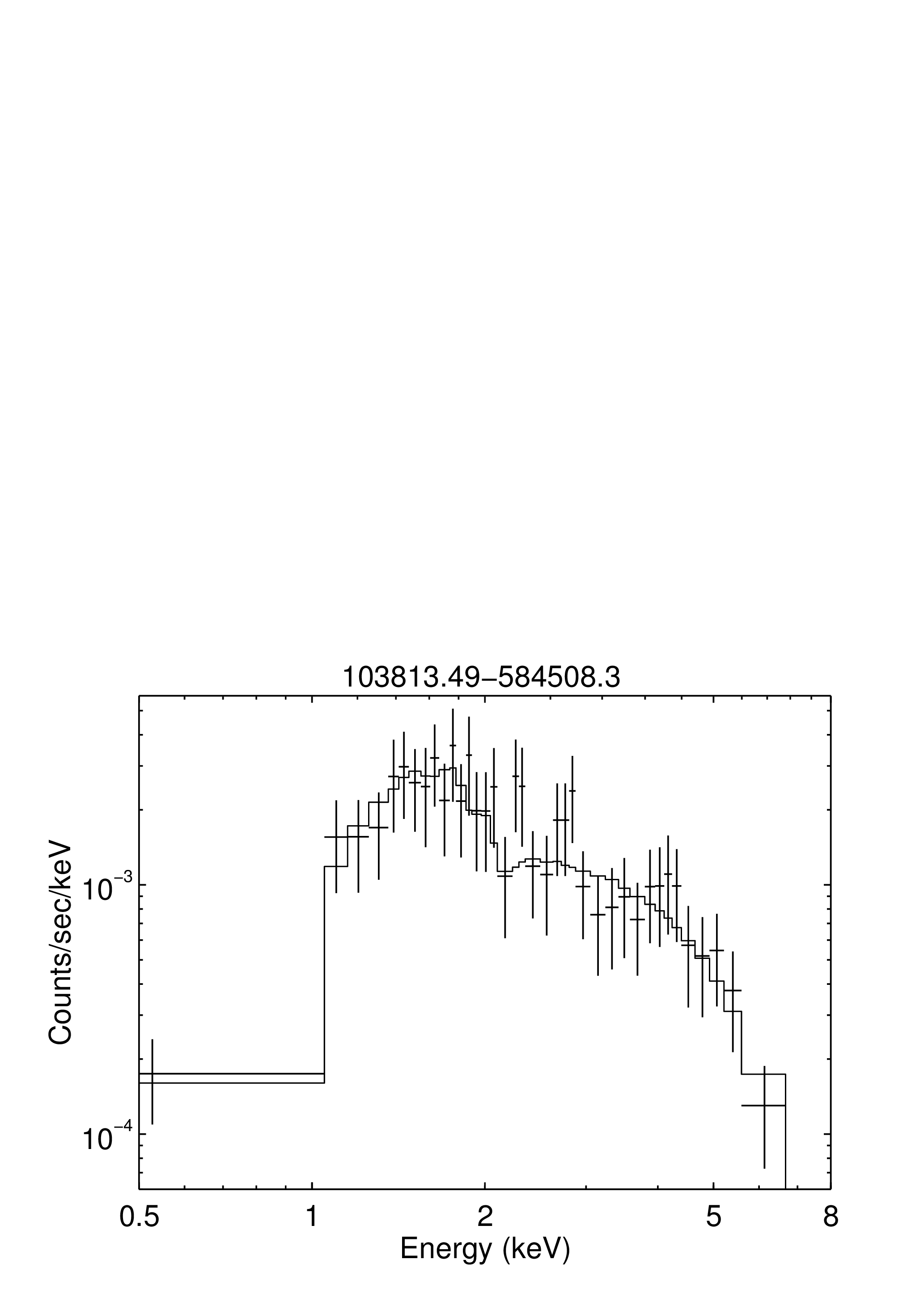}}
   \caption{\textit{Chandra} X-ray spectra and best-fit models
of four bright X-ray sources in our Gum~31 observation.
The crosses show the measured spectra, the solid lines show the best-fit
models. 
}
              \label{fig:spec}%
    \end{figure*}

\begin{table*}
\caption{Spectral fitting results for the 22 X-ray sources with more than 100 counts. 
The table lists for each source the quality of the
fit expressed by the $\chi^2_n$, the absorbing hydrogen column density $N_{\rm H}$,
the plasma temperature $ kT $, the normalization (which is the emission measure of the plasma
 scaled by the distance $D$,
i.e.,~norm~$= \int n_e n_H \,d V\, \large/ \,[ 4 \pi D^2]$\, ),
and finally the intrinsic (i.e.,~extinction corrected) X-ray luminosity in the total ($0.5-8$~keV) band,
$L_{\rm X,tc}$. For the brightest source J103722.27$-$583723.0, which corresponds to the
O6.5 star HD~92\,206~A, we list the parameters for a single plasma temperature model fit
and for
two- and three-temperature model fits; these fits are discussed in 5.2.1.}
\label{tab:spectra}
\centering
\begin{tabular}{l c  c c c c}
\hline\hline
\noalign{\smallskip}
 Source     &$\chi^2_n$ & $N_{\rm H}$  &  $ kT $   & norm     & $\log (L_{\rm X,tc})$ \\
          & & $[10^{22}\,{\rm cm}^{-2}]$  &  [keV]    &  $[10^{-19}\,{\rm cm}^{-5}]$           & [erg/s] \\
\noalign{\smallskip}
\hline
\noalign{\smallskip}
J103652.59$-$583627.8  & 0.25  &$0.33  \pm 0.33$& $1.15 \pm  0.41$& $4.08 \pm  0.29$& $31.06 \pm 0.16$\\
J103653.22$-$583633.2  & 0.89  &$1.80  \pm 0.51$& $1.99 \pm  0.47$& $8.99 \pm  3.06$& $31.24 \pm 0.23$\\
J103653.40$-$583737.5  & 0.83  &$0.56  \pm 0.30$& $0.96 \pm  0.35$& $6.73 \pm  3.92$& $31.16 \pm 0.33$\\
J103700.41$-$583829.2  & 0.14  &$0.13  \pm 0.14$& $2.73 \pm  0.71$& $1.77 \pm  0.42$& $30.96 \pm 0.15$\\
J103701.61$-$583512.0  & 0.54  &$0.27  \pm 0.12$& $7.29 \pm  3.53$& $4.68 \pm  0.65$& $31.56 \pm 0.12$\\
J103706.34$-$583633.7  & 0.67  &$0.26  \pm 0.20$& $3.03 \pm  1.25$& $9.83 \pm  3.27$& $31.73 \pm 0.14$\\
J103716.31$-$583617.4  & 0.74  &$0.25  \pm 0.19$& $4.19 \pm  3.78$& $2.93 \pm  0.89$& $31.26 \pm 0.18$\\
J103718.63$-$583741.9  & 0.55  &$0.26  \pm 0.17$& $0.57 \pm  0.14$& $3.68 \pm  1.87$& $30.98 \pm 0.29$\\
J103719.02$-$583749.1  & 0.57  &$0.43  \pm 0.27$& $1.37 \pm  0.52$& $3.02 \pm  1.31$& $30.90 \pm 0.22$\\
J103722.56$-$583655.6  & 0.67  &$0.43  \pm 0.15$& $2.45 \pm  0.54$& $3.27 \pm  0.67$& $31.13 \pm 0.13$\\
J103722.95$-$583722.9  & 0.53  &$0.20  \pm 0.09$& $0.68 \pm  0.08$& $4.70 \pm  1.52$& $31.21 \pm 0.13$ \\
J103730.26$-$585004.4  & 0.60  &$0.00  \pm 0.14$& $2.51 \pm  1.67$& $2.27 \pm  0.55$& $31.11 \pm 0.09$\\
J103736.29$-$583929.4  & 0.43  &$0.84  \pm 0.42$& $5.32 \pm  11.1$& $3.23 \pm  1.56$& $31.28 \pm 0.20$\\
J103745.86$-$584024.1  & 0.25  &$0.00  \pm 0.12$& $1.95 \pm  0.76$& $1.41 \pm  0.34$& $30.89 \pm 0.11$\\
J103752.94$-$584407.2  & 0.10  &$0.39  \pm 0.15$& $2.40 \pm  0.90$& $3.19 \pm  0.79$& $31.13 \pm 0.19$\\
J103754.06$-$584642.3  & 0.42  &$1.48  \pm 0.47$& $3.86 \pm  3.31$& $5.66 \pm  2.42$& $31.35 \pm 0.26$\\
J103758.57$-$584647.9  & 0.55  &$3.62  \pm 0.33$& $2.61 \pm  3.12$& $10.0 \pm  9.84$& $31.24 \pm 0.35$\\
J103807.71$-$584313.0  & 0.83  &$0.00  \pm 0.02$& $1.00 \pm  0.08$& $2.12 \pm  0.23$& $31.08 \pm 0.05$\\
J103813.49$-$584508.3  & 0.43  &$0.95  \pm 0.15$& $6.59 \pm  1.56$& $10.6 \pm  1.28$& $31.80 \pm 0.09$\\
J103814.39$-$584658.3  & 0.44  &$0.35  \pm 0.15$& $0.79 \pm  0.14$& $3.65 \pm  1.24$& $30.73 \pm 0.18$\\
J103816.23$-$584235.8  & 0.26  &$0.09  \pm 0.11$& $0.76 \pm  0.12$& $3.00 \pm  0.99$& $31.08 \pm 0.20$\\
\hline  \noalign{\smallskip}
J103722.27$-$583723.0  & 0.81  &$0.31 \pm  0.06$& $0.73  \pm 0.06$& $22.7 \pm 3.79$ & $31.78 \pm 0.12$\\ 
 (1 temperature fit)   & & & & & \\
\noalign{\smallskip}
J103722.27$-$583723.0  & 0.56  &$0.32 \pm  0.14$& $0.38 \pm  0.05$& $28.7 \pm 15.0$ & \\
 (2 temperature fit)                   &         &      $+$        &$2.32 \pm  1.14$& $6.55 \pm  2.11$& $31.85 \pm 0.21$\\
 \noalign{\smallskip}
J103722.27$-$583723.0  & 0.49  & $0.46 \pm  0.46$&$0.20 \pm  0.08$& $60.2\pm 56.9$ & \\
  (3 temperature fit)                  &         &      $+$        &$0.54 \pm  0.54$& $24.9 \pm 5.61$   & \\
                    &         &      $+$        &$5.17 \pm  5.17$& $3.21 \pm 1.70$& $31.88 \pm 0.24$ \\
\hline                                   
\end{tabular}
\end{table*}

For the 
22 sources in our sample with more than 100 net counts we performed a 
spectral fitting analysis
using the {\it SHERPA} v4.5 software in CIAO \citep[see][]{Freeman01}.
{\it SHERPA} includes the plasma emission and absorption
 models of \textit{XSpec} version 11.3. 
The background-subtracted spectra were first grouped into bins containing
at least 10 counts, and the fits were then performed using the 
$\chi^2$ statistic with the \citet{Gehrels86} variance 
function\footnote{The standard deviation is calculated with the formula
$\sigma = 1 + \sqrt{N + 0.75}$, which is thought to be 
more realistic than assuming a Poisson distribution,
if the number of counts per bin is low. We note that for spectra with
10 counts per bin, the expected value for the reduced chi-square statistic
with the Gehrels variance 
is $E[\chi^2_n] = 0.74$.},
which is the {\it SHERPA} default.

We used models with one or more thermal plasma
{\it VAPEC} components, and the
{\it TBABS} model to describe the effect of
extinction by interstellar and/or circumstellar material (as
parameterized by the hydrogen column density $N_{H}$).
The elemental abundances for the plasma model were fixed at the
values\footnote{The adopted abundances, relative to the solar photospheric
abundances given by \citet{ag89}, are:
C = 0.45, N = 0.788, O = 0.426, Ne = 0.832, Mg = 0.263, Al = 0.5, 
Si = 0.309, S = 0.417, Ar = 0.55, Ca = 0.195, Fe = 0.195, Ni = 0.195.}
that were found by \citet{Guedel07} to be typical
for young stellar objects.

Most spectra could be well reproduced by one-temperature models.
The only case where models with more than one temperature components
produced significantly better fits was the O star HD~92\,206~A, which has 920
source counts (this X-ray spectrum is discussed in more detail
in Sect.~\ref{individual.sect}).
A few selected examples of the spectral fits are shown in Figure \ref{fig:spec},
while the resulting best fit spectral parameters for all sources 
are reported in Table \ref{tab:spectra}.  We also list there
 the intrinsic (i.e.,~extinction corrected)
X-ray luminosities $L_{X,tc}$ for the total ($0.5-8$~keV) band,
as derived from the spectral fit parameters.

The range of X-ray luminosities spans from
$\log \left( L_{X,tc}\,[{\rm erg/s}] \right) = 30.73$ to 31.88.
The hydrogen column densities derived in the fits are mostly around
$N_{\rm H} \sim 0.3 \times 10^{22}\,{\rm cm}^{-2}$, corresponding to
visual absorptions of $A_{V} \sim 1.5$~mag.
Only a few sources show column densities above
$10^{22}\,{\rm cm}^{-2}$ ($A_{V} \ga 5$~mag). The highest
extinction of  
$N_{\rm H} = 3.6 \times 10^{22}\,{\rm cm}^{-2}$ ($A_{V} \sim 18$~mag)
is found for the source J103758.57$-$584647.9, a faint infrared source
in the dense cloud associated to the cluster G286.38--0.26.
These values are in good agreement with the cloud column densities
derived from our \textit{Herschel} far-infrared data
\citep[see][and Fig.~4 in Ohlendorf et al.~2013]{Preibisch12}.

The derived plasma temperatures 
range from $\approx 0.57$~keV ($\approx 7$~MK) for J103718.63$-$583741.9,
up to $\sim 7.29$~keV ($\approx 85$~MK) for J103701.61$-$583512.0.
The later source is a faint, optically visible star near the northwestern edge
of the \ion{H}{ii} region,   
which showed a strong flare during our \textit{Chandra} observation
(see lightcurve in Fig.~\ref{fig:lightcurves}).

The derived  plasma temperatures and X-ray luminosities are 
in the typical ranges found for YSOs in other star forming regions
\citep[see, e.g.,][]{Preibisch_coup_orig}. 
A detailed discussion of individual interesting 
sources will be given in Sect.~\ref{individual.sect}.

\subsection{X-ray luminosity estimates for fainter sources}
\label{ssec:xphot}
Most of the X-ray sources have less
than 100 source counts, which is a practical limit
for detailed X-ray spectral analysis.
For these fainter sources,
we used the \textit{XPHOT}
software\footnote{www.astro.psu.edu/users/gkosta/XPHOT/},
developed by \citet{get10}, to derive an estimate of the intrinsic
(i.e.,~extinction corrected) X-ray luminosity of the X-ray sources.
\textit{XPHOT} is based on a non-parametric method for the calculation
of fluxes and absorbing X-ray column densities of weak
X-ray sources. X-ray extinction and intrinsic flux are estimated from
the comparison of the apparent median energy of the
source photons and apparent source flux with those of high signal-to-noise
spectra that were simulated using
spectral models characteristic of much brighter sources of similar class
previously studied in detail. This method requires $\ge 5$ net counts
per source  and can thus be applied to 286 of our 679 sources.
For the remaining sources, the ``energy flux''
determined by \textit{ACIS Extract} can be used to compute a rough estimate
of the X-ray luminosity \citep[see discussion in][for more details]{CCCP-catalog}.
For those 22 source for which a spectral fit was performed, 
a comparison shows that the XPHOT and the spectral fit 
X-ray luminosity estimates agree within a factor of two
in most cases.


\subsection{Spatial distribution of the X-ray sources}

   \begin{figure}
   \centering
\includegraphics[width=9.0cm]{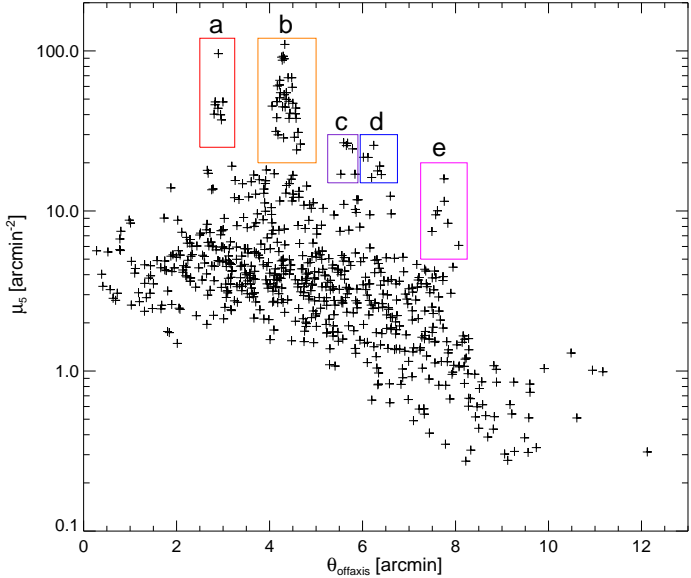}
   \caption{
Nearest neighbor surface density $\mu_5$ at the location 
of each X-ray source plotted against the offaxis-angle. The general 
decrease in density with increasing offaxis-angle is related to
instrumental effects. The boxes enclose the members of the
identified clusterings.}
              \label{fig:nn-density}%
    \end{figure}

We performed a nearest neighbor analysis \citep[see][]{CH85} to
obtain a quantitative characterization of the spatial distribution of X-ray
sources.
The nearest neighbor technique  allows 
statistically significant overdensities to be identified in an objective way
and is widely used in studies of the clustering properties of star forming
regions \citep[see, e.g.,][]{Gutermuth09}.
In the present study, we used the distance to the fifth nearest neighbor
to calculate the surface density estimator $\mu_5$
at the position of each source.
For the interpretation of the resulting densities,
we have to take the spatial variations
of the detection sensitivity over the ACIS field-of-view
into account:  sensitivity is highest in the center, and
decreases (due to factors  such as mirror vignetting and the
increasing width of the point-spread function) towards
the edges of the field-of-view.
We therefore plot in Fig.~\ref{fig:nn-density} 
the surface density estimator $\mu_5$ as a function of the
offaxis-angle.
The general trend of decreasing
source density with increasing offaxis-angle above about $5'$
can be clearly seen.
Clusters of X-ray sources can be identified as spatially 
confined groups of sources 
for which the local surface density clearly exceeds 
the values found at other locations in the image at similar
offaxis-angles.

We identify five different clusterings, \textsf{a} to \textsf{e},
 in Fig.~\ref{fig:nn-density}.
Clustering \textsf{a} (8 members) is a rather small, but quite dense
group of stars, which is located about $1.5'$ southeast of the
cluster NGC~3324 at the coordinates  
 $\alpha ({\rm J2000}) = 10^{\rm h}\,37^{\rm m}\,31^{\rm s}$,
$\delta ({\rm J2000}) = -58\degr\,38'\,30''$.
Clustering \textsf{b} (36 members) corresponds to the stellar
cluster NGC~3324.
Clustering \textsf{c} (6 members) corresponds to a small, but dense
group of stars, about $1'$  northwest of the
cluster NGC~3324, at the coordinates
 $\alpha ({\rm J2000}) = 10^{\rm h}\,37^{\rm m}\,07^{\rm s}$,
$\delta ({\rm J2000}) = -58\degr\,37'\,00''$.
The group \textsf{d} (7 members) corresponds to the
cluster G286.38--0.26 in the southern parts of the ACIS image.
The group \textsf{e} (7 members) corresponds to the
a group of stars at the northwestern 
inner rim of the \ion{H}{ii} region, at the coordinates
 $\alpha ({\rm J2000}) = 10^{\rm h}\,36^{\rm m}\,59^{\rm s}$,
$\delta ({\rm J2000}) = -58\degr\,35'\,23''$.

\subsection{Expected contamination of the X-ray source sample}
\label{ssec:contamination}

As in any X-ray observation,
there will be some degree of contamination
by galactic field stars in the fore- and background as well as extragalactic sources.
To quantify the expected level of this contamination,
we use here  the results from the 
\textit{Chandra} Carina Complex Project 
that observed the central parts of the CNC with 
very similar exposure times ($\approx 60 - 80$~ks)
as our Gum~31 pointing. Since our Gum~31 pointing is very close
to the sky region covered by the CCCP, the contamination levels 
should be quite similar.

The detailed simulations of contaminating X-ray-emitting populations by
\citet{Getman11} predicted $\sim 5000$ contaminating 
sources\footnote{The specific predictions were $\sim 1800$ foreground stars,
$\sim 900$ background stars, and $\sim 2300$ AGN.}
 in the CCCP field. This would imply that about
35\% of all X-ray sources in the CCCP field are contaminants.
The classification study of
\cite{CCCP-classification}, which considered the
X-ray, optical, and infrared properties of the sources
(that differ for the different contaminant classes), found that
75\% of CCCP sources can be classified as young stars in the Carina nebula, 
while 11\% were classified as contaminants, and 14\% remained unclassified 
(e.g.,~due to incomplete and/or ambiguous data).
This leads to a slightly lower estimate of the 
contamination rate, between $\ge 11\%$ and $\sim 25\%$.

Considering these two different estimates, we therefore 
assume a contamination rate of about $25-30\%$ for the
X-ray sources in our Gum~31 pointing, and can thus expect
$\sim 180$ contaminants and $\sim 500$ young stars in our
sample of 679 X-ray sources.

\section{Optical and Infrared data}

Here we describe the data sets which we used for
the identification and characterization of the X-ray sources.

\subsection{Optical images}

In order to search for possible optical counterparts of the
 X-ray sources, we first inspected
optical images from the Digitized Sky Survey as well as
$V$- and $R$-band images
obtained with the Wide Field Imager on the MPG/ESO 2.2~m telescope
(which are deeper and provide better angular resolution than the DSS images).
This search revealed possible optical counterparts to
371 (54.6\%) of the X-ray sources.

We also inspected the available \textit{Hubble Space Telescope} (HST)
 archive images
in the Gum~31 region. A mosaic consisting of a
$\approx 3.25 ' \times 6.5'$ wide box covering
the western part of the ionization front and a (partly overlapping)
$\approx 3.25' \times 3.25'$ box covering the northern half of the
cluster NGC~3324 had been obtained with ACS/WFC in the
context of the 10th anniversary of the 
Hubble Heritage Project\footnote{see
{\tt http://hubblesite.org/newscenter/archive/releases/\\
2008/34/}.
A description of these data and the available high-level products 
sets can be found at
{\tt http://archive.stsci.edu/prepds/carina/}.}.
These HST images cover a total area of about 30 square-arcminutes,
i.e.,~only about 10\% of the field-of-view of our \textit{Chandra}
observation.
123 X-ray sources are located in the area covered by the HST images, and
84 of these (i.e.,~68.3\%) have an optical counterpart.

In total, 382 of the 679 X-ray sources have an optical counterpart in the
DSS, WFI and/or HST images.
It is immediately clear that a reasonably complete
 optical characterization of low-mass stars in the
Gum~31 would require much deeper optical data than available:
assuming  an age of about 3~Myr and a typical extinction of $A_V = 2$~mag,
stars with masses of $[\,1.0, \,0.5, \,0.1\,]\,M_\odot$ are predicted to
have visual magnitudes of $V= [\,19.3, \,21.0, \,26.3\,]$
according to the pre-main-sequence stellar models of
\citet{Siess00}. This shows that the majority 
of the low-mass ($\leq 1.0\,M_\odot$)
stellar population will thus remain undetectable in the available
optical images.

The near-infrared regime is much better suited to
detect and characterize the young low-mass stars: for
the above mentioned stellar masses, ages, and extinctions,
the expected magnitudes are $J= [\,15.7, \,16.7, \,18.0\,]$.
These stars will be rather easily detectable in even relatively
moderately deep near-infrared images (although not in the 2MASS data).

\subsection{VISTA near-infrared survey data}

We have recently used the
4\,m Visible and Infrared Survey Telescope for Astronomy \citep[VISTA;][]{Emerson06}
at the European Southern Observatory under program number 088.C-0117(A)
to obtain a deep and very wide ($\approx 2.3\degr \times 2.9 \degr$)
near-infrared survey that covers the full spatial extent of
 the CNC, including the entire area of Gum~31.
The results of this survey are very well
suited for an identification and characterization of the
counterparts of the X-ray sources in Gum~31. 
While a detailed description of our infrared survey 
and the
resulting source catalog will be given in Preibisch et al.~(in prep.), the
main aspects can be summarized as follows:
we obtained deep VISTA images 
in the near-infrared $J-$, $H-$, and $K_s-$bands during
March 2012.
VISTA's infrared camera VIRCAM \citep{Dalton06} consists of an array of sixteen individual 
$2048 \times 2048$ pixel infrared detectors
with a nominal mean pixel size of 0.339\arcsec\ on the sky.
In order to account for the gaps between the individual detectors,
six individual exposures with corresponding
offset in x- and y-direction are combined to yield a so-called `tile',
that covers an area of $1.2\degr \times 1.5\degr$ without gaps.
The entire area of our Gum~31 \textit{Chandra} pointing lies within the
northwestern tile of our $2 \times 2$ tile mosaic VISTA survey.
These data were processed by the VISTA data flow system 
\citep[see][]{Irwin04} at the 
Cambridge Astronomy Survey Unit, which provided catalogs 
of all detected point-like sources.
The photometric calibration was performed by using stars
in the 2MASS Point Source Catalog, which are
present in large
numbers (several thousands) in each VISTA tile, as will be described in
detail in Preibisch et al.~(2014, in prep.).
The final photometric uncertainties, characterized
by the standard deviations between the calibrated VISTA magnitudes
and 2MASS magnitudes for stars in the magnitude intervals
of $(J, H, K_s) \sim [12 \dots 13]$  are in the range $\sigma = [0.04\,\dots\,0.05]$~mag 
for all three bands.
For the bright stars,
that are in or close to the non-linear or saturated regime in at least
one of the three VISTA bands
(i.e.,~$J \le 12.0$, $H \le 11.75$, or $K_s \le 11.5$),
photometry from the 2MASS Point Source Catalog was used.

Our final VISTA catalog contains more than four million individual sources,
3\,951\,580 of which are detected in at least two of the three NIR bands.
The formal $5 \sigma$ detection limit for pointlike sources  in the Gum~31 area
(calculated from the measured skynoise in the corresponding tiles)
is $J \approx 20.4$, $H \approx 19.4$, and $K_s \approx 18.8$.
Typical values for the completeness limit across the field  are
$J_{\rm compl} \approx 18.5$, $H_{\rm compl} \approx 18$, 
and $K_{s,\, \rm compl} \approx 17.5$; nearly all objects brighter than these limits 
are S/N~$\ge 10$ detections.
Comparing these numbers to the above mentioned pre-main sequence models,
we find that our VISTA catalog should be complete for moderately-obscured 
($A_V \la 5$~mag) young ($\la 3$~Myr) stars 
down to masses of $\approx 0.1\,M_\odot$.
The number of cataloged NIR sources in the area of our Gum~31
\textit{Chandra} observation is about 46\,000.

As any large astronomical source catalog, the VISTA catalog is not 100\% perfect.
Especially in regions with strong diffuse nebulosity and near the
extended point-spread-functions of very bright stars, 
the detection efficiency is limited and some point-like sources
that are clearly visible in the images were not detected by the
VISTA data processing pipeline and are thus missing from the
catalog. Although this problem concerns only a very small fraction
(about $1-2\%$) of the total survey area, a few infrared counterparts
of X-ray sources in the Gum~31 area are  missing from the catalog.
However, as described below, these cases could be easily identified
and solved 
by visual inspection of the original VISTA images.

\subsection{\textit{Spitzer}  mid-infrared data}

For the identification of mid-infrared counterparts of the
X-ray sources, we used our \textit{Spitzer} point source
catalog that we created from 
all available IRAC archive data of the CNC as described in 
\citet{Ohlendorf14}.
The estimated completeness limits of our IRAC catalog are
$\approx 1.5$\,mJy, $\approx 0.7$\,mJy, $\approx 1.2$\,mJy, and $\approx 1.6$\,mJy for 
the IRAC 1, 2, 3, and 4 bands.
With these limits, the photospheres of young ($\la 3$~Myr) stars
with masses down to $\approx 0.5\,M_\odot$ can be detected
(at least) in the IRAC~1 and 2 bands,
for moderate extinction ($A_V \la 5$~mag).
There are about 4300 \textit{Spitzer} catalog sources in the area of the
\textit{Chandra} observation of Gum~31.

\subsection{\textit{Herschel} far-infrared data}

Maps of the far-infrared emission in the CNC
were obtained in December 2010 in our \textit{Herschel} 
Open Time project. We used the PACS/SPIRE parallel mode to
produce maps in the 70, 160, 250, 350, and $500\,\mu$m
band for a more than 5 square-degree wide area that covers the
full spatial extent of the CNC and includes the Gum~31 area.
A full description of these observations and the
data processing can be found in \cite{Preibisch12}.
The angular resolution of the \textit{Herschel} maps ranges
from $5''$ in the $70\,\mu$m map to $25''$ in the
$500\,\mu$m map.

Our \textit{Herschel} images revealed 642 reliably detected point-like
sources, which have been analyzed in \cite{Gaczkowski13}.
As described there in detail, the sensitivity limit of these data
maps allows the detection of 
pre-stellar cores with cloud masses of  $\ga 1\,M_\odot$ and
Class~0 protostars with stellar masses of $\ga 1\,M_\odot$.
Young stellar objects in later evolutionary phases are only
detected as long as they still have rather massive circumstellar
envelopes and/or disks; 
solar-mass objects will be detected only during their class~I phase,
but not in later phases.
Intermediate mass objects with disks may
also be detected during their class II phases.

There are 35 \textit{Herschel} point-like sources in the field of
our \textit{Chandra} observation of Gum~31.

\section{Infrared counterparts of the X-ray sources \label{ircp.sect}}

\subsection{Source matching procedure}

In order to identify infrared counterparts of the X-ray 
sources in our VISTA and \textit{Spitzer} catalogs, 
we performed a two-step process. The first step is an automatic 
matching that identifies counterparts solely based on the
X-ray and infrared source coordinates. The second step included a detailed
individual inspection of all X-ray source positions in the
infrared images  and also utilized the available
a-priori information to resolve problematic cases such as 
multiple possible matches.

For the first step, the automatic source matching,
we employed the method described by 
\citet{CCCP-catalog} as implemented in the IDL 
tool\footnote{see\, {\tt www2.astro.psu.edu/xray/docs/TARA/TARA\_users\_guide/node11.html}} 
{\tt match\_xy.pro}. 
The maximum acceptable separation between an X-ray source and a counterpart 
is based on the individual source position errors assuming Gaussian 
distributions, scaled so that $\sim ∼99\%$ of true associations 
should be identified as matches.
The X-ray source position errors determined by AE
range from $0.04''$ to $1.45''$, with a mean value of $0.46''$.
For the VISTA sources, we assumed position uncertainties of
$0.1''$, while for the \textit{Spitzer} sources, we used the 
magnitude-dependent
position uncertainties up to $0.5''$ as derived by the
MOPEX software \citep[see][for details]{Ohlendorf13}.

In the first stage, the algorithm tests the hypothesis that a possible 
pair of sources from the two catalogs is spatially coincident.
The most significant match of each master source is referred to as its 
``Primary Match''; any other
significant matches are ``Secondary Matches''.
The second stage of the algorithm resolves possible 
many-to-one and one-to-many relationships between the X-ray 
catalog and the infrared catalogs.
Clear one-to-one relationships are classified as ``successful primary matches'',
while in cases where, e.g., two X-ray sources are significantly close to
a single infrared source, the less significant primary match
 is labeled as ``failed''.
This finally provides a reasonable one-to-one set of matches.

\subsection{\textit{Chandra} -- VISTA matching results}

The matching procedure yielded 
481 successful primary VISTA matches to the 679 \textit{Chandra}
sources.
There was no case of failed primary matches, but
for 16  X-ray sources one or more successful secondary matches
were identified.
These 16 X-ray sources have more than one possible infrared 
counterpart within the matching box, and it is not guaranteed
that the closest match is always the correct physical counterpart.
 Since the surface density of infrared sources in our deep
VISTA images is rather high\footnote{The average surface density
of VISTA catalog sources in the field of our
\textit{Chandra} pointing is 136 objects per square-arcminute.
Since the typical 99\% match region has a area of about
3~square-arcseconds, the expected probability of random associations
with unrelated infrared sources is about 10\%.},
it is possible that
physically unrelated (e.g., background) infrared sources
may appear in the matching region just by chance
and produce a ``false match'', which even could degrade the 
true infrared counterpart to a secondary match.

We can use a-priory information about the infrared properties
of the X-ray detected young stars to identify possible cases where
this problem occurs.
Since we know from the  X-ray detection limit that most of the
X-ray detected objects should be young stars with
 masses of $\ga 0.5\,M_\odot$,
these stars should typically be relatively bright NIR sources.
Therefore, all cases where an X-ray source has an unexpectedly faint 
primary match
and a considerably brighter secondary match, deserve special attention.
We found four cases where the primary match was very faint
($J \ge 18$, or $H \ge 17.5$, or $K_s \ge 17.2$) 
and a brighter secondary match was present.
For these cases,
we replaced the original
very faint primary match by this brighter secondary match.

Our final detailed inspection of the VISTA images
revealed several cases, where an X-ray source for which  the
procedure {\tt match\_xy.pro} did not report a match,
has a rather bright infrared source  
very close to the X-ray source position,
but just outside the matching region. 
It is very likely that
some of these cases are ''false negative'' rejections
of physical matches. A quantitative estimate of the expected number
of ''false negative'' rejections can be made as follows:
In theory, the use of
 99\% probability boundaries for the
match region should lead
to about 1\% false negative rejections of true matches, i.e.,~about
7 cases in our X-ray sample.
In reality,
this number will be somewhat higher, since the true
positional errors will not perfectly follow a Gaussian distribution,
but usually have somewhat wider ``wings''. 
This estimate agrees quite well with the number of 11
observed bright infrared sources that are located within
$\le 1.0''$ of the X-ray source position but just outside the
formal matching region.
We therefore decided to add these 11 cases to our catalog of
infrared counterparts.
In all other cases, where the angular distance between an  X-ray source
and a star in the VISTA images is larger than $1.0''$, 
we did {\em not} add matches ``by hand'',
in order to keep the matching as objective as possible.
Although relaxing the matching limits
would yield a slightly higher rate of matches, 
it would also increase the number
of unphysical chance associations. We prefer here a clean sample over a
marginally larger, but less clean sample.

Finally, our visual inspection of the X-ray source positions
on the VISTA images revealed 8 cases where a clear match
is visually evident,
but the infrared source was missing from the VISTA catalog.
For these stars, we performed individual aperture photometry in
the VISTA tile images and added them to our counterpart list.

In total and as the final result of our matching procedures,
we have VISTA NIR counterparts for 500 of the 679 X-ray sources;
the corresponding matching rate is 73.6\%.
For 458 of these sources, VISTA photometry is available in all three bands.
Table~3 \onltab{3}{}  lists the near-infrared magnitudes of the
X-ray sources.

We note that these matching results are very well consistent with the
expectation about the number of young stars ($\sim 500$) and 
contaminants ($\sim 180$) in the
sample of X-ray sources as discussed in Sect.~2.6,
since the population of contaminants should be dominated by 
extragalactic objects,
and almost all X-ray detected AGN are expected to be so faint in the
NIR regime ($J \ga 20$) that they should remain
undetected in our VISTA images.

\subsection{\textit{Chandra} -- \textit{Spitzer} matching results}

The matching of the X-ray source list 
with our \textit{Spitzer} IRAC point source catalog
with the {\tt match\_xy} procedure produced 272 successful primary matches.
Since the positional uncertainties of mid-infrared sources in the
\textit{Spitzer} images are higher than those of the
NIR sources in the VISTA images, the maximum allowed matching radii used for the
\textit{Chandra} -- \textit{Spitzer} matching are larger than those
used for the \textit{Chandra} -- VISTA matching.
Therefore, an additional step is necessary to make sure to get consistent
\textit{Chandra} -- VISTA -- \textit{Spitzer} matches.
For all reported \textit{Spitzer} matches to \textit{Chandra}  sources
we checked whether the \textit{Spitzer} source position is 
consistent with a NIR source in the VISTA images.
There were 14 cases, where a
\textit{Spitzer} source that was classified as a match to an X-ray source
could be clearly identified with a VISTA sources that was \textit{not} a match 
of the \textit{Chandra} source.
In these cases, we removed the \textit{Spitzer} match.  

The final detailed visual inspection of all X-ray source positions in the
\textit{Spitzer} images revealed ten sources where a clear infrared counterpart
can be seen, but is missing from the \textit{Spitzer} point 
source photometry catalog.

The whole identification procedure finally yields
268 \textit{Spitzer} counterparts to the  679 X-ray sources;
the corresponding matching rate is 39.5\%.
For 215 of these sources photometry is available in the $3.6\,\mu$m band,
for 258 in the $4.5\,\mu$m band, for 98 in the $5.6\,\mu$m band, and
for 46 in the $8.0\,\mu$m band.
The IRAC magnitudes of the X-ray sources are listed
in Table~3.

\subsection{\textit{Chandra} -- \textit{Herschel} matching results}

Because of the relatively moderate angular resolution of the \textit{Herschel}
images and the often weak contrast between point-like sources
and the generally strong surrounding cloud emission,
the positional uncertainties for the
\textit{Herschel} point-like sources are considerably larger than those
of the NIR and MIR sources.
Therefore, we used a search radius of $5''$ to look for possible
X-ray detections of \textit{Herschel} sources.

Five X-ray sources are
located within $5''$ of a \textit{Herschel} source.
In all five cases, the X-ray source
has a clear match with a NIR point source visible in the VISTA images. 
Therefore, these objects are no pre-stellar or very young protostellar (class~0)
objects, but more evolved YSOs of class~I or class~II.


\subsection{X-ray and infrared sources in the clusters NGC~3324 and G286.38--0.26}

   \begin{figure*}  \sidecaption
\includegraphics[width=11.9cm]{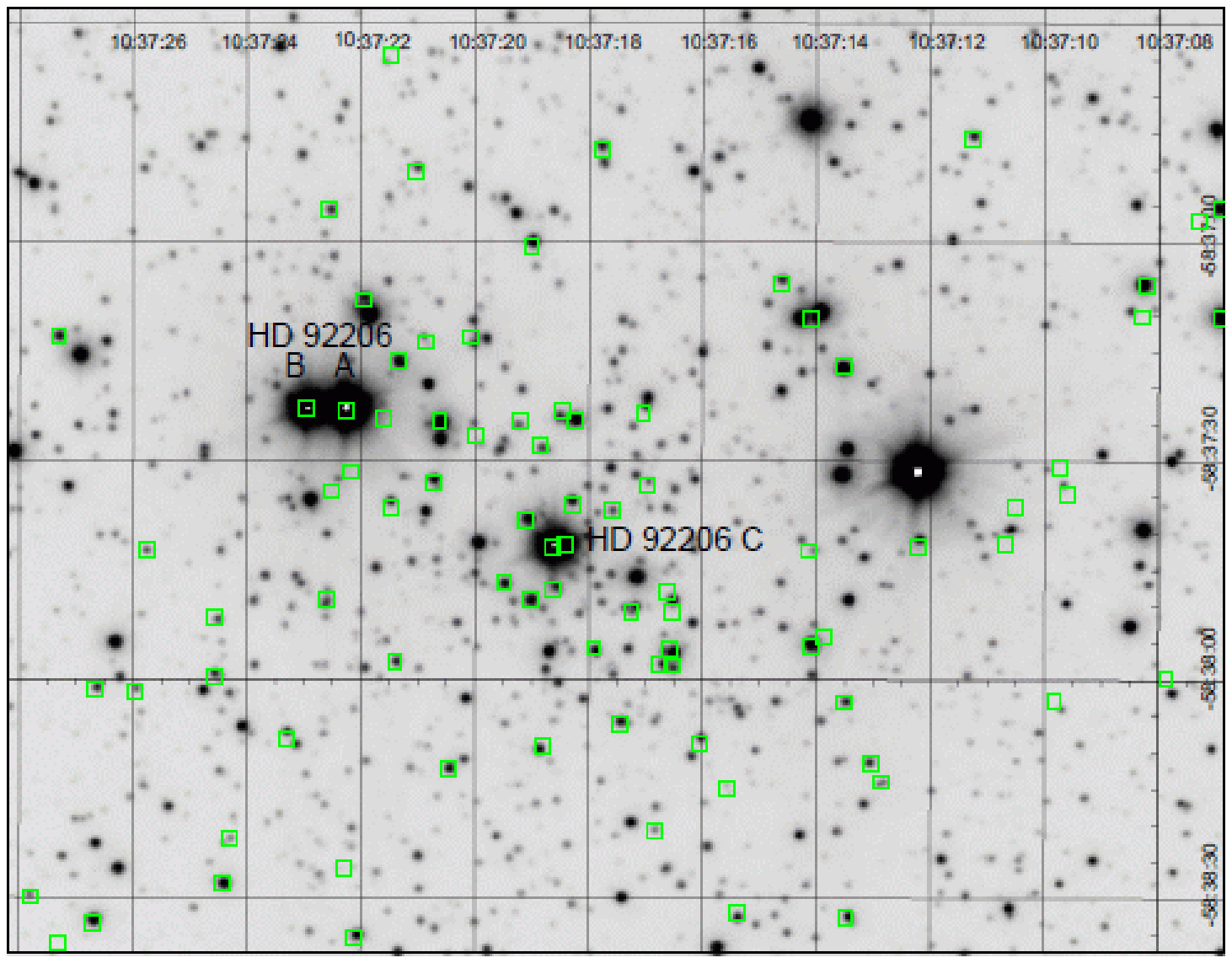}\\
 \includegraphics[width=11.9cm]{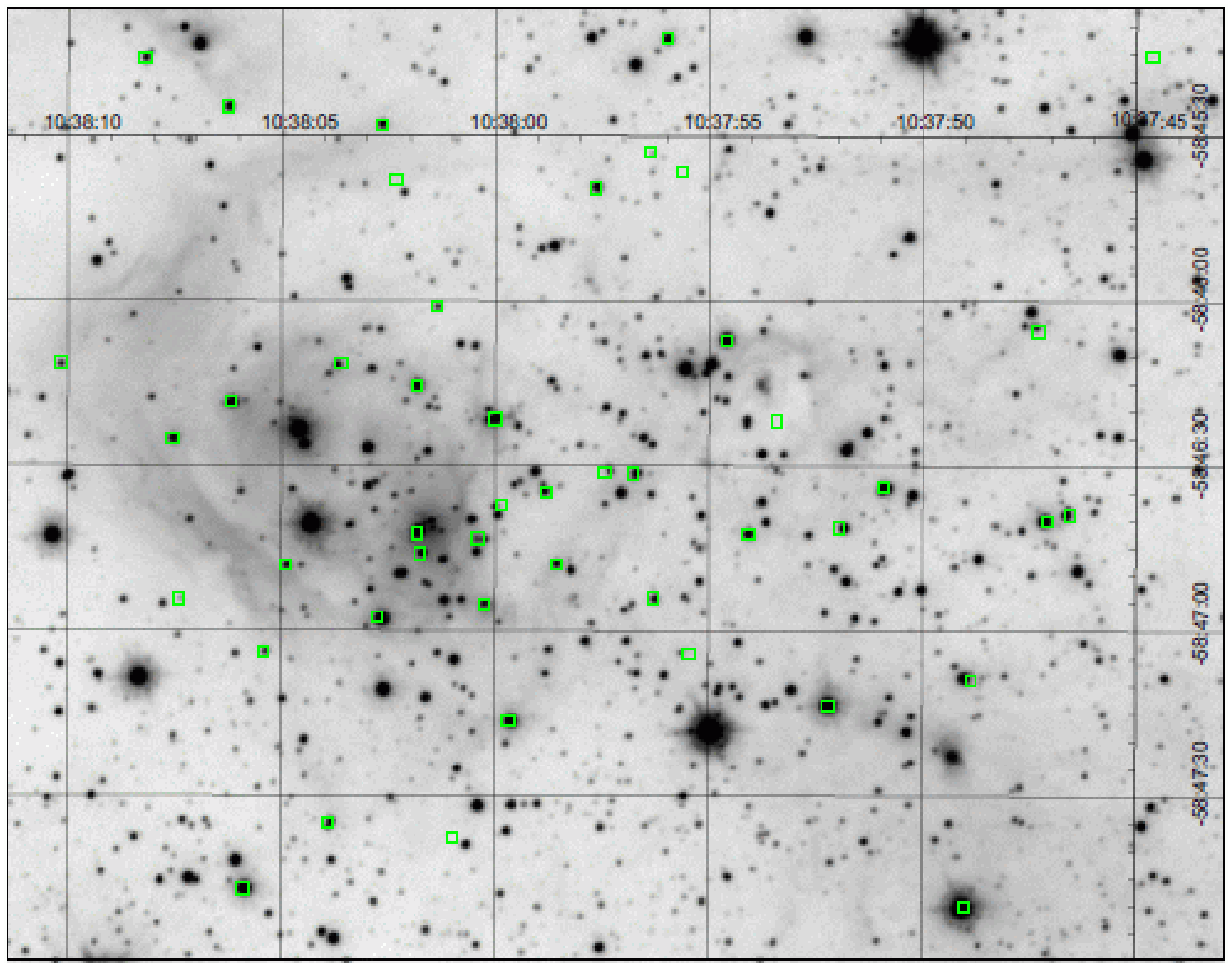}
  \caption{Top: Negative representation of the VISTA $H$-band image 
of the cluster NGC~3324 (top) and 
the VISTA $K_s$-band image 
of the cluster G286.38--0.26 (bottom).
The positions of the X-ray sources are shown by the green boxes.
We note that the X-ray source position uncertainties
 are in most cases
considerably smaller than the size of the box symbols; the rather large $2''$
box size was chosen for a clarity of display. 
  }
              \label{n3324-g286.fig}%
    \end{figure*}

The stellar cluster NGC~3324 harbors the highest and densest concentration of
X-ray sources (see Fig.~\ref{n3324-g286.fig}).
It contains 82 X-ray sources in a $2.5' \times 2.5'$ region.
Among these are the three known O-type stars in the Gum~31 region.
The VISTA image shows several hundred faint stars, many of which are probably 
low-mass cluster members with X-ray luminosities below our detection limit.
A more detailed discussion of the X-ray detected population will be given in Sect.~\ref{pop.sect}.
Neither the VISTA or \textit{Spitzer}, nor the \textit{Herschel} images
show significant amounts of cloud emission in or around this cluster. This implies
that no active star formation is going on in this region, and the young stars
have already largely dispersed their natal clouds.
\bigskip

The stellar cluster G286.38--0.26 is the second most significant spatial concentration of
young stars in the area.
We find 42 X-ray sources in a $ 3.8' \times  2'$  region.
In contrast to NGC~3324, this cluster is still associated to large amounts
of cloud material, which is prominently seen in the \textit{Herschel} and
\textit{Spitzer} images and is traced by the diffuse nebulosity seen in the
above VISTA image. 
The cloud column-densities derived from the \textit{Herschel} data
for this region \citep[see][]{Preibisch12,Ohlendorf13} ranges from about 
$N_{\rm H} \approx 1 \times 10^{22}\,{\rm cm}^{-2}$ ($A_V \approx 5$~mag) to
$N_{\rm H} \approx 4 \times 10^{22}\,{\rm cm}^{-2}$ ($A_V \approx 20$~mag).
Embedded in these clouds are several bright protostellar
infrared sources (some of which are discussed in the next Section). 
However, there is also a number of optically visible stars
which show rather low extinction and no infrared excesses.
Thus, the cluster contains a mix of very young objects,  ongoing
star formation activity, and more evolved, older stars.
A multi-wavelength image comparison of this cluster 
can be seen in Fig.~12 in \citet{Ohlendorf13}.

\section{X-ray properties of the brightest infrared and optical objects \label{individual.sect}}

The properties of the particularly interesting objects,
that warrant a detailed look, are described in this section.

\subsection{Embedded Infrared Sources}

\paragraph{J103801.84$-$584642.4:} This source is
identified with the infrared source MSX6C~G286.3747-00.2630 in the
 cluster G286.38--0.26.  This object  is very faint in 
optical images, very bright in the near-infrared, and
one of the brightest mid-infrared 
sources in this area (especially in the $24\,\mu$m MIPS image).
These properties suggest it 
to be a very young, embedded YSO with large amounts of
circumstellar material. 
The X-ray lightcurve shows no strong variability.
The emission is relatively hard (median photon energy 2.33~keV), and 
XPHOT estimates a column density of $N_{\rm H} \sim 1.5 \times 10^{22}\,{\rm cm}^{-2}$
(corresponding to a visual extinction of $A_V \sim 8$~mag).
The
X-ray luminosity estimated with XPHOT is $\log \left( L_{\rm X} {\rm [erg/s]} \right) \approx 31.4$.
According to the general relation between X-ray luminosity and stellar mass for young stellar
objects derived in the \textit{Chandra} Orion Ultradeep Project \citep[see][]{Preibisch_coup_orig}, 
steady X-ray emission
at such a level suggest a stellar mass in the range of $\sim 2-3\,M_\odot$.

\paragraph{J103800.40$-$584643.3:}  This
 is an optically invisible source in the G286.38--0.26 cluster. It is invisible
in our VISTA $J$- and $H$-band images (implying $J \ga 21$, $H \ga 20$),
but appears as a faint  $K_s$-band source 
with $K_s \sim 16$.
It is quite bright in the IRAC images, but undetected in the MIPS image.
The \textit{Herschel} far-infrared maps show a dense cloud clump
with a hydrogen column density of $N_{\rm H} \approx 3 \times 10^{22}\,{\rm cm}^{-2}$
($A_V \approx 15$~mag)
near its position, suggesting that this source is a
particularly deeply embedded YSO.
This source showed a strong flare during our \textit{Chandra} observation
and has a very hard median photon energy of 5.4~keV.
XPHOT estimates a very high column density of $N_{\rm H} \sim 4 \times 10^{23}\,{\rm cm}^{-2}$
(corresponding to a visual extinction of $A_V \sim 200$~mag) and a very high
X-ray luminosity of $\log \left( L_{\rm X} {\rm [erg/s]} \right) \approx 32.4$ for this object.

\paragraph{J103806.57$-$584000.4:} This \textit{Chandra} source corresponds to
an optically invisible infrared source 
with very red near-infrared colors ($J-H = 2.92$,
$H -K_s = 1.60$), which is
located at the tip of a prominent pillar at the eastern edge of the
\ion{H}{ii} region. An image of this object and the pillar and the
near- to far-infrared spectral energy distribution can be seen in 
\citet{Ohlendorf13}, where it is listed as J103806.6$-$584002.
It is classified as a class~I object
with a stellar mass of $M_\ast \approx 1.7\,M_\odot$ that is surrounded by a
circumstellar disk and a massive envelope and has a
total luminosity of $L \approx 40\,L_\odot$.
The XPHOT estimate of the column density of $N_{\rm H} \sim 2 \times 10^{22}\,{\rm cm}^{-2}$
(corresponding to a visual extinction of $A_V \sim 10$~mag)
is consistent with these infrared results. 
The XPHOT estimate of its  X-ray luminosity is 
$\log \left(L_{\rm X} {\rm [erg/s]} \right) \approx 30.7$, suggesting a fractional X-ray luminosity of
$\log \left(L_{\rm X} / L_{\rm bol} \right) \approx -2.9$.

\subsection{Optically bright stars}

\subsubsection{The early-type (O,B,A) stars}

   \begin{figure} 
   \centering
\parbox{8cm}{\includegraphics[width=8.0cm,bb = 31 26 540 419, clip]{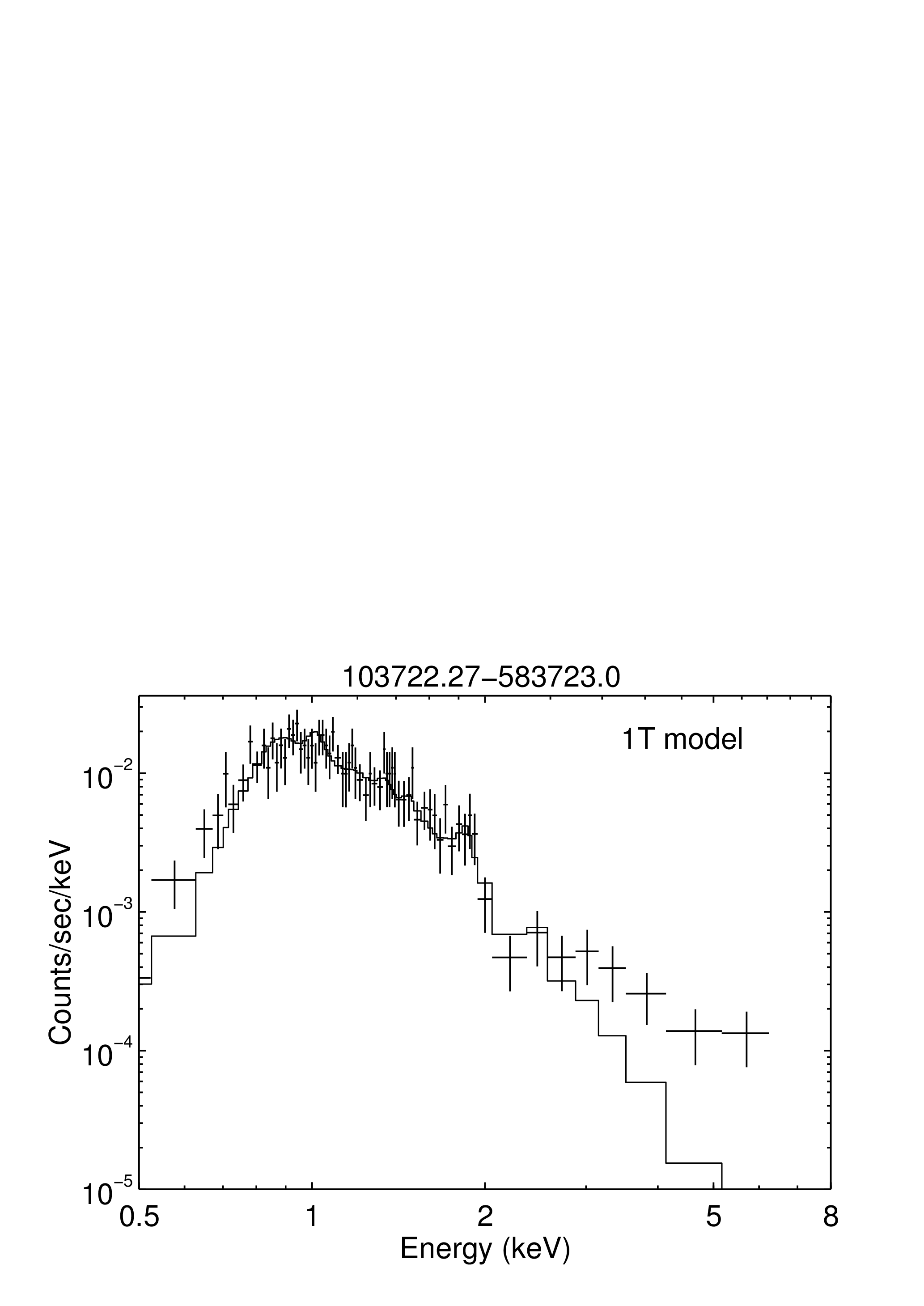}\\
 \includegraphics[width=8.0cm,bb = 31 26 540 398, clip]{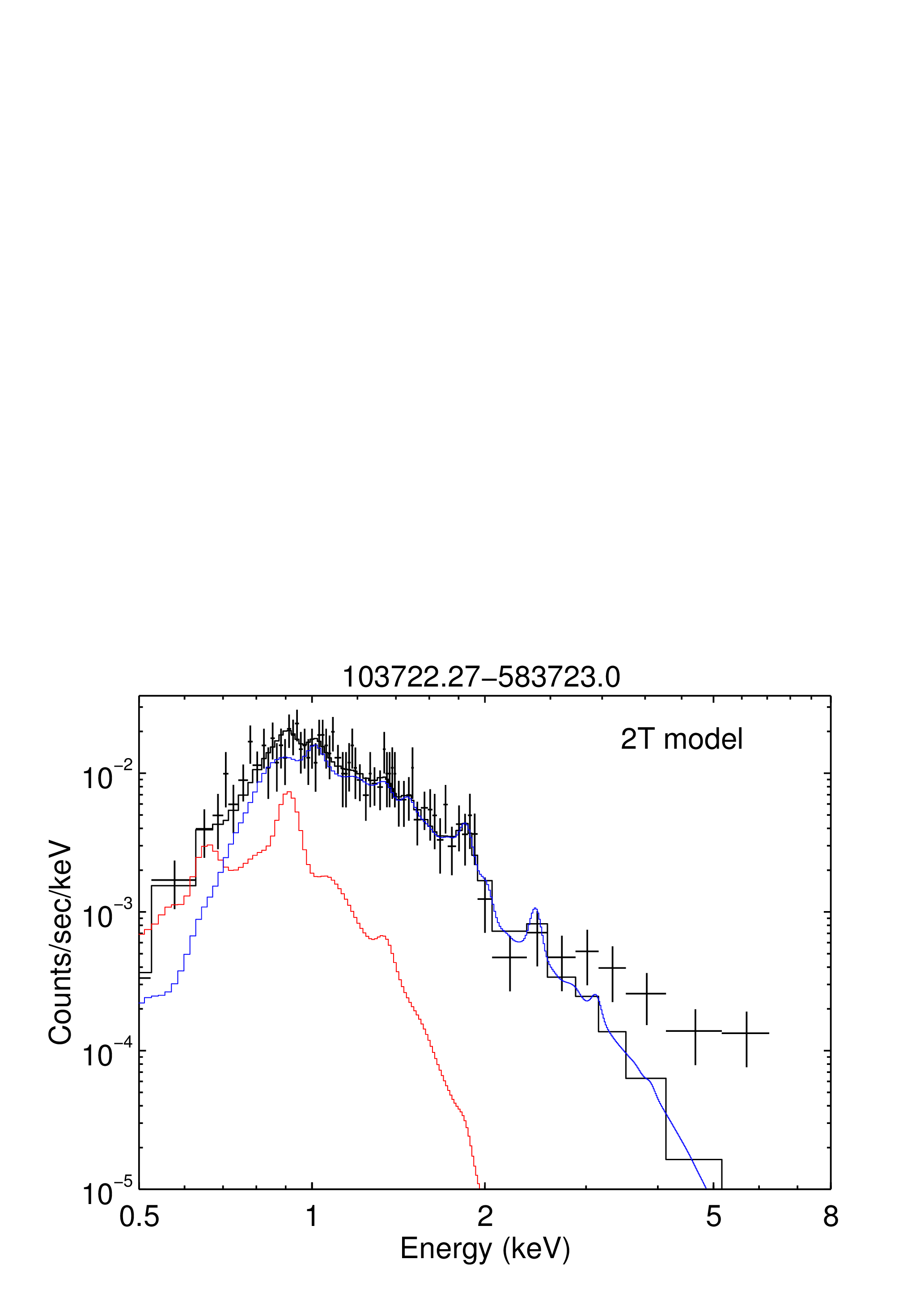}\\
 \includegraphics[width=8.0cm,bb = 31 26 540 400, clip]{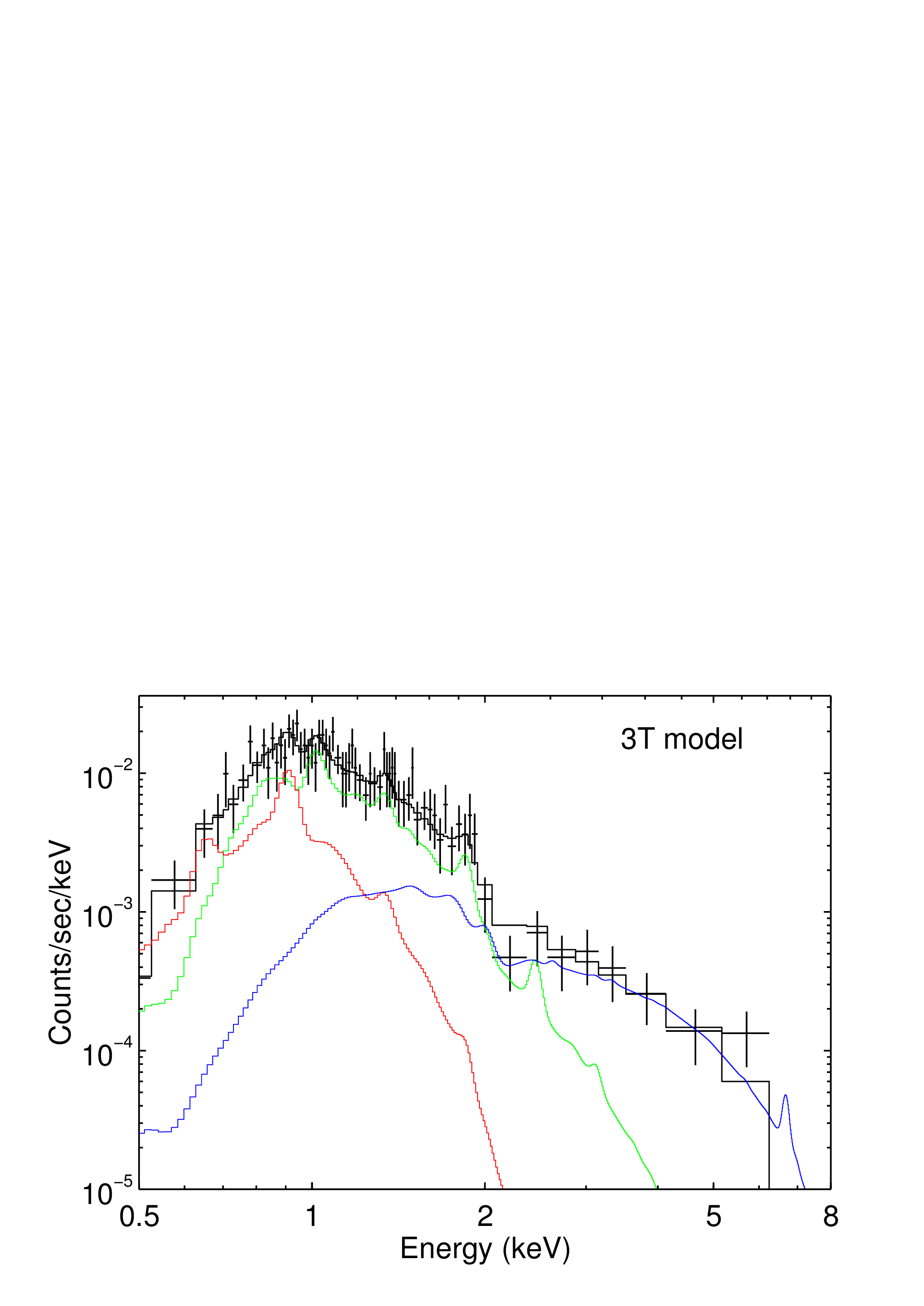}\\
 \includegraphics[width=8.4cm]{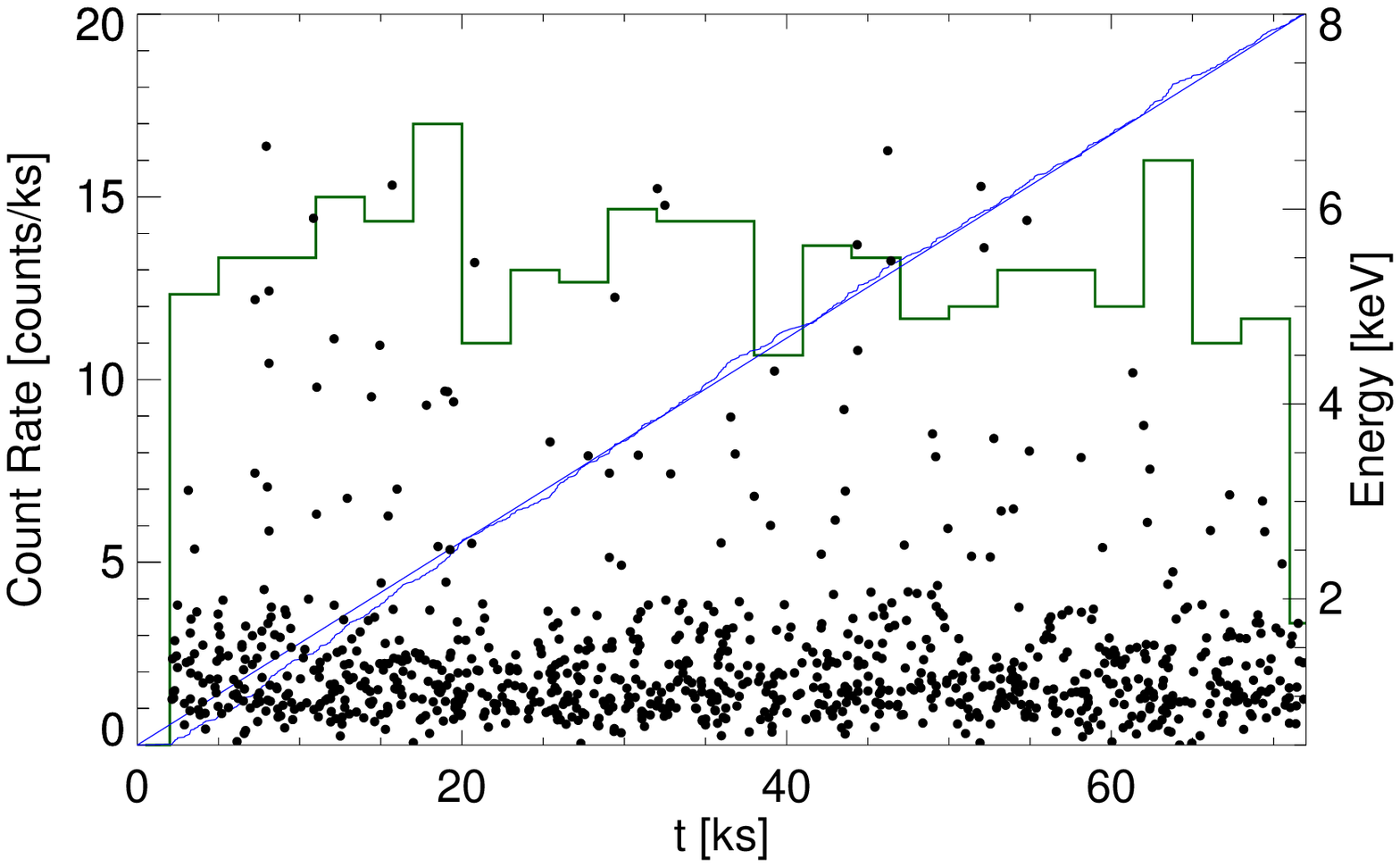}}
   \caption{X-ray spectral fits and lightcurve of the O6.5 star
HD~92\,206 A.
  }
              \label{HD92206-fits.fig}%
    \end{figure}

The three known O-type stars in Gum~31 are the components
A, B, and C of HD~92\,206, located in the central cluster NGC~3324.
They are the most luminous and massive stars in the region.
All three are clearly detected as X-ray sources, and our 
analysis reveals two additional, yet unreported components.

\paragraph{HD~92\,206~A:}
The optically brightest component HD~92\,206~A  is of spectral type O6.5 V.
It can be clearly identified with the brightest X-ray source (J103722.27$-$583723.0) 
in our \textit{Chandra}
observation. The 920 source counts allow a rather detailed
analysis of the X-ray spectrum. 
The fit with a single-temperature spectral model 
(Fig.~\ref{HD92206-fits.fig}, top) leaves
systematic excesses in the soft as well the hard parts of the 
spectrum.
Although a two-temperature spectral model yields a statistically
acceptable fit with temperature components of
$kT_1 = 0.38 \pm 0.1$~keV and $kT_2 = 2.32 \pm  1.14$~keV,
there is still a systematic excess in the observed spectrum at high
photon energies (see Fig.~\ref{HD92206-fits.fig}).
We therefore considered a three-temperature model, which yields
a better fit with temperature components of
$kT_1 = 0.20 \pm 0.08$~keV, $kT_2 = 0.54 \pm 0.54$~keV, 
and $kT_3 = 5.17 \pm 5.17$~keV. 
 Although the
temperatures are not very tightly constrained, 
this model reproduces the shape of the spectrum very well
(see Fig.~\ref{HD92206-fits.fig}).
The ratio of the corresponding
emission measures is $60.2 : 24.9 : 3.21$,
clearly showing
that the X-ray spectrum is dominated by rather soft
emission with plasma temperatures of
$\la 6 \times 10^6$~K, as typical for O-type stars. 
The X-ray luminosity of $\log \left( L_{\rm X}\,[{\rm erg/sec}] \right) = 31.88 \pm  0.24$
is also in the typical range for late O-type stars.
Assuming that the bolometric luminosity of HD~92\,206~A is
 $\log \left(L_{\rm bol}/L_\odot \right) = 5.23$ according to its spectral type
and the models of \citet{Martins05}, 
we find a fractional X-ray luminosity of
$\log \left( L_{\rm X}/L_{\rm bol} \right) \approx -6.9$.
This is very close to the canonical value of fractional 
X-ray luminosities for O-type stars in general and also in the
Carina nebula:
\citet{Gagne11} found that the great majority of O 
stars in the Carina nebula have $\log \left( L_{\rm X}/L_{\rm bol}\right)$
 ratios between 
$-6.8$ and $-8.0$, with a mean value of $-7.23$.

   \begin{figure}
   \centering
\includegraphics[width=8cm]{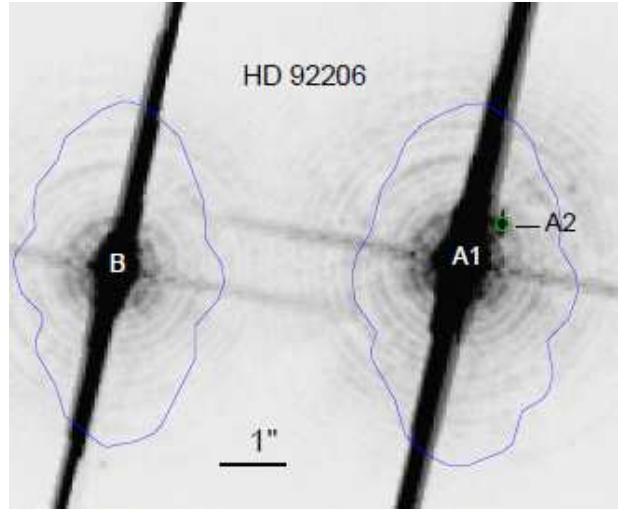}
   \caption{Optical Hubble Space Telescope image of HD~92\,206 A and B
in the filter F658N; north is up and east is to the left.
The blue polygons show the X-ray extraction regions,
which enclose 90\% of the PSF 
(note that the Chandra PSF is elongated and enlarged at the 
rather large off-axis angle ($4.3'$) of the object).
The faint companion of HD~92\,206 A  can be seen about $0.8''$
northwest and is marked as ``A2''.
  }
              \label{HD92206A.fig}%
    \end{figure}

The hard 5.17~keV
component (corresponding to a plasma temperature of 
about $60 \times 10^6$~K),
is however, quite unusual for an O-type star.
One possibility is that this unexpected hard X-ray emission
originates not from the O-star but from another nearby
object, such as a coronally active late type companion.
We therefore performed a detailed inspection of the available optical and infrared 
images of HD~92\,206~A to explore this possibility. 
An archival Hubble Space Telescope image, obtained with the WFC
camera of ACS in the filter F658N, is shown in Fig.~\ref{HD92206A.fig}.
This image shows a faint point source
at the position RA $= 10^{\rm h}\,37^{\rm m}\,22.206^{\rm s}$, 
DEC $=-58\degr\,37'\,22.26''$,
i.e.,~at an angular offset of $0.83''$ from the O-star,
corresponding to a projected physical separation of
$\approx 1900$~AU.
We will designate the companion as HD~92\,206~A2, and the
primary star as HD~92\,206~A1.
As can be seen in Fig.~\ref{HD92206A.fig}, 
the X-ray source position is in very good agreement with the
location of the bright O-star primary HD~92\,206~A1, clearly showing
that most of the observed X-ray emission must predominantly
come from HD~92\,206~A1, not from the companion A2. However,
the location
of the companion is close enough that its X-ray photons
are clearly included in the extraction region that is centered
on HD~92\,206~A1.

In order to obtain an estimate of the stellar properties of the companion,
we performed aperture photometry in another HST ACS/WFC image that 
was obtained in the medium-band filter F550M (unfortunately, not
data obtained through broad-band filters are available).
This yielded a magnitude of $m_{\rm F550M} \approx 18.1 \pm 0.5$,
where the rather large uncertainty is dominated by the choice of the
sky region, since the object is located in the outer parts of the
point-spread-function of the much brighter star HD~92\,206~A1.
According to the PMS stellar models of \citet{Siess00},
the derived magnitude is consistent with the assumption
that the companion A2 is a young ($\sim 1$~Myr old) star with a 
mass of about $0.5 - 1\,M_\odot$, if the extinction is not much higher
than $A_V \la 3$~mag. The companion could thus well be a young
late-type star, which produces coronal X-ray emission, what
might explain the excess of hard X-ray emission seen in the spectrum
of HD~92\,206~A.
We note that the X-ray lightcurve  shows no significant variability;
therefore, the hard X-ray emission of the system HD~92\,206~A1/A2
is \textit{not} related to coronal flaring activity.

\paragraph{HD~92\,206~B}
(= CD~$-57\,3380$~B; spectral type O6.5 V) 
corresponds to X-ray source J03722.95-583722.9 and
yielded 226 source counts in our \textit{Chandra} observation.
A fit to the X-ray spectrum with a two-temperature model yields
plasma components 
with $kT_1 = 0.22 \pm 0.05$~keV and $kT_2 = 0.64 \pm 0.05$~keV
and gives an X-ray luminosity of 
$\log \left( L_{\rm X}\,[{\rm erg/sec}] \right) = 31.20 \pm  0.15$.
Assuming again a bolometric luminosity of 
 $\log (L_{\rm bol}/L_\odot) = 5.23$ according to its spectral type
and the models of \citet{Martins05},
we find a fractional X-ray luminosity of
$\log \left( L_{\rm X}/L_{\rm bol} \right) \approx -7.6$ for
this star.
The lightcurve shows no significant variability.
These properties are in the typical range as reported for
the O-type stars in the Carina nebula by \citet{Gagne11}.

   \begin{figure}
   \centering
 \includegraphics[width=8cm]{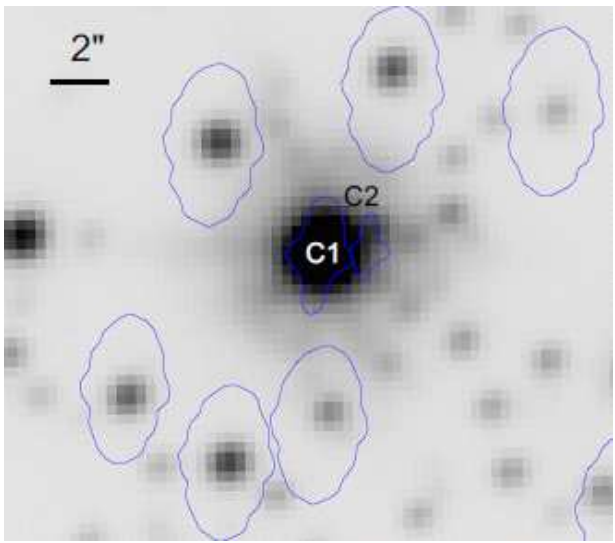}
 \caption{VISTA $K_s$-band image of the region around HD~92\,206 C;
north is up and east is to the left.
The blue polygons show the individual X-ray extraction regions.
The faint companion of HD~92\,206 C  can be seen about $1.7''$
northwest and is marked as ``C2''.
The extraction region for the source C2 was automatically shrunken 
to avoid overlap with the extraction region of source C1, 
and thus encloses only 40\% of the PSF.
We note that the Chandra PSF is elongated and enlarged at the
rather large off-axis angle ($4.3'$) of theses objects.
  }
              \label{HD92206C.fig}%
    \end{figure}

\paragraph{HD~92\,206~C}
(= CD~$-57\,3378$) is the third known O-type star
in NGC~3324. A spectral type of O8.5Vp is listed in the
SIMBAD catalog. However, the spectroscopic monitoring study of
\citet{Campillay07} found HD~92206~C to be a double-lined spectroscopic
binary with a period of 2.02~days. 
They derived spectral types of O7.5 V + B0 V for the two
components and determined a mass-ratio of about 0.7 from the 
radial velocity curve analysis.

HD~92\,206~C is
clearly detected by \textit{Chandra} as source J103718.63$-$583741.9.
It yielded 119 source counts, 
and our spectral fit 
gives a plasma temperature of  $kT = 0.57 \pm 0.14$~keV and
an X-ray luminosity of
$\log \left( L_{\rm X}\,[{\rm erg/sec}] \right) = 30.98 \pm  0.29$.
Assuming a bolometric luminosity of
 $\log \left( L_{\rm bol}/L_\odot \right) = 5.05$ (according to the
O7.5 spectral type of the primary
and the models of \citet{Martins05}),
we find a fractional X-ray luminosity of
$\log \left( L_{\rm X}/L_{\rm bol} \right) \approx -7.3$ for
this star. The lightcurve shows at most marginal indications for
variability.
These values are again in the typical range observed for
late O-type stars.

Interestingly, the \textit{Chandra} image revealed another 
X-ray source, J103718.41$-$583741.6,
 at an angular separation of just $1.7''$ from HD~92\,206 C. 
Our detailed inspection of the VISTA images reveals
a marginally resolved companion near the position of the
second X-ray source and thus confirms the presence
of two distinct sources (see Fig.~\ref{HD92206C.fig}).
We designate the secondary component as HD~92\,206 C2.
Unfortunately, the small angular separation from the very
bright O-type primary
star C1 did not allow us to obtain reliable photometric
flux estimates from the VISTA data. 
All we can say is that the companion is considerably fainter
in the near-infrared than the main component.
In the available HST images,
this stellar system is located just outside the field-of-view
and can thus not be analyzed.

The extraction of the X-ray source J103718.41$-$583741.6 had to be done
with a reduced aperture size in order to avoid overlap with the
nearby source J103718.63$-$583741.9 (=HD~92\,206~C);
the extraction region contains only 40\% of the PSF and
yielded just 16 net counts, too few for a detailed
spectral analysis. The median photon energy is 1.19~keV, and the
estimated X-ray luminosity derived with XPHOT is
$\log \left( L_{\rm X}\,[{\rm erg/sec}] \right) \approx  30.57$.
The fact that the companion is considerably less bright
in the X-ray as well as in the near-infrared regime,
suggests that it is probably a late-type star.

\paragraph{The known B- and A-type stars} in Gum~31,
HD~303\,080 (spectral type B),
HD~92\,145 (spectral type A2), HD~92\,207 (spectral type A0Iae),
and HD~303\,094 (spectral type A2),
are not detected as X-ray sources.
This is not surprising, since no X-ray emission is 
expected from stars in the spectral range $\sim$B2 to $\sim$A9
\citep[see, e.g., discussion in][]{Stelzer05}.

\subsubsection{The Algol-type eclipsing binaries KU Car and DT Car}

   \begin{figure}
   \centering
 \parbox{7.5cm}{\includegraphics[width=7.5cm]{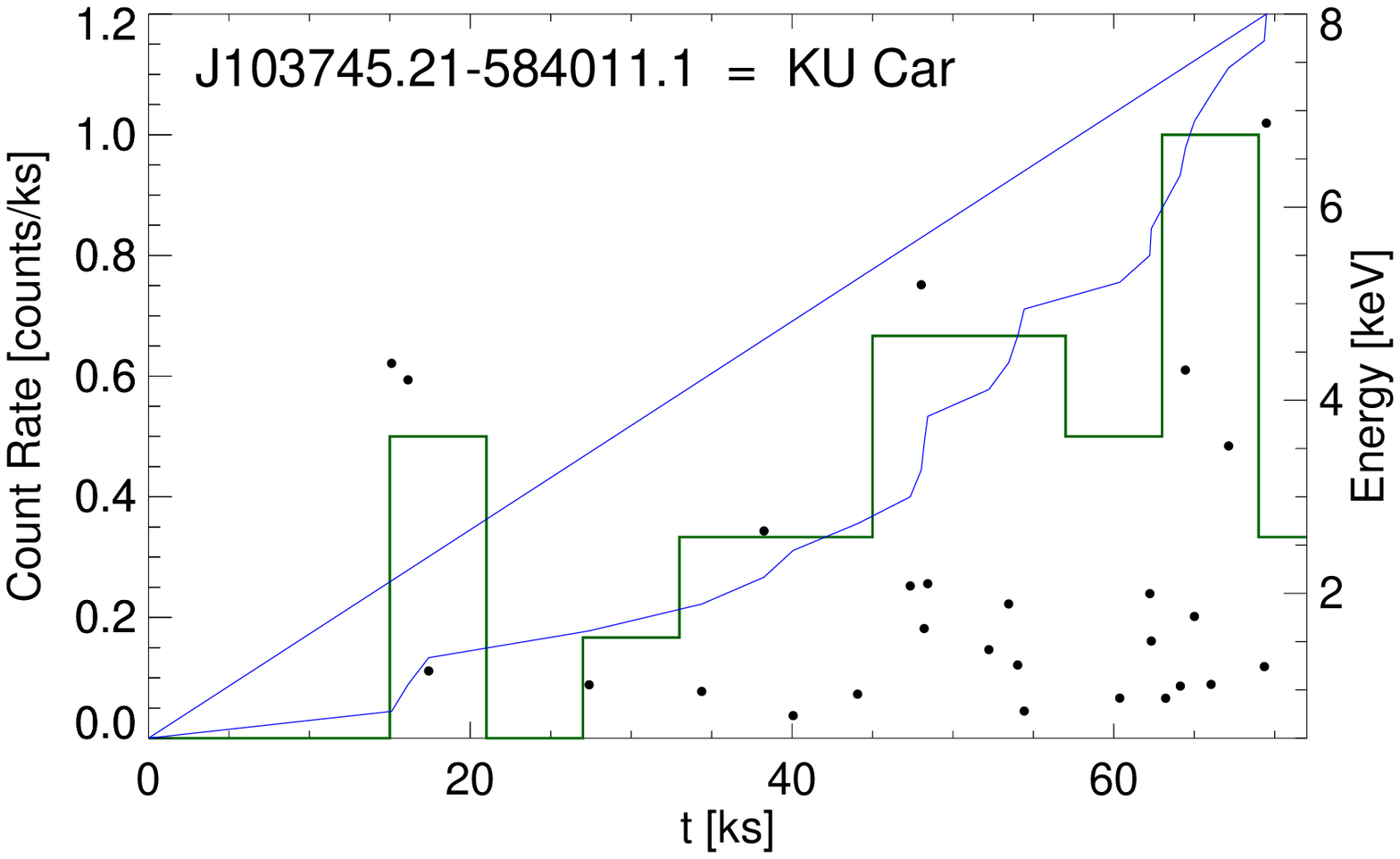}\vspace{2mm}

 \includegraphics[width=7.5cm]{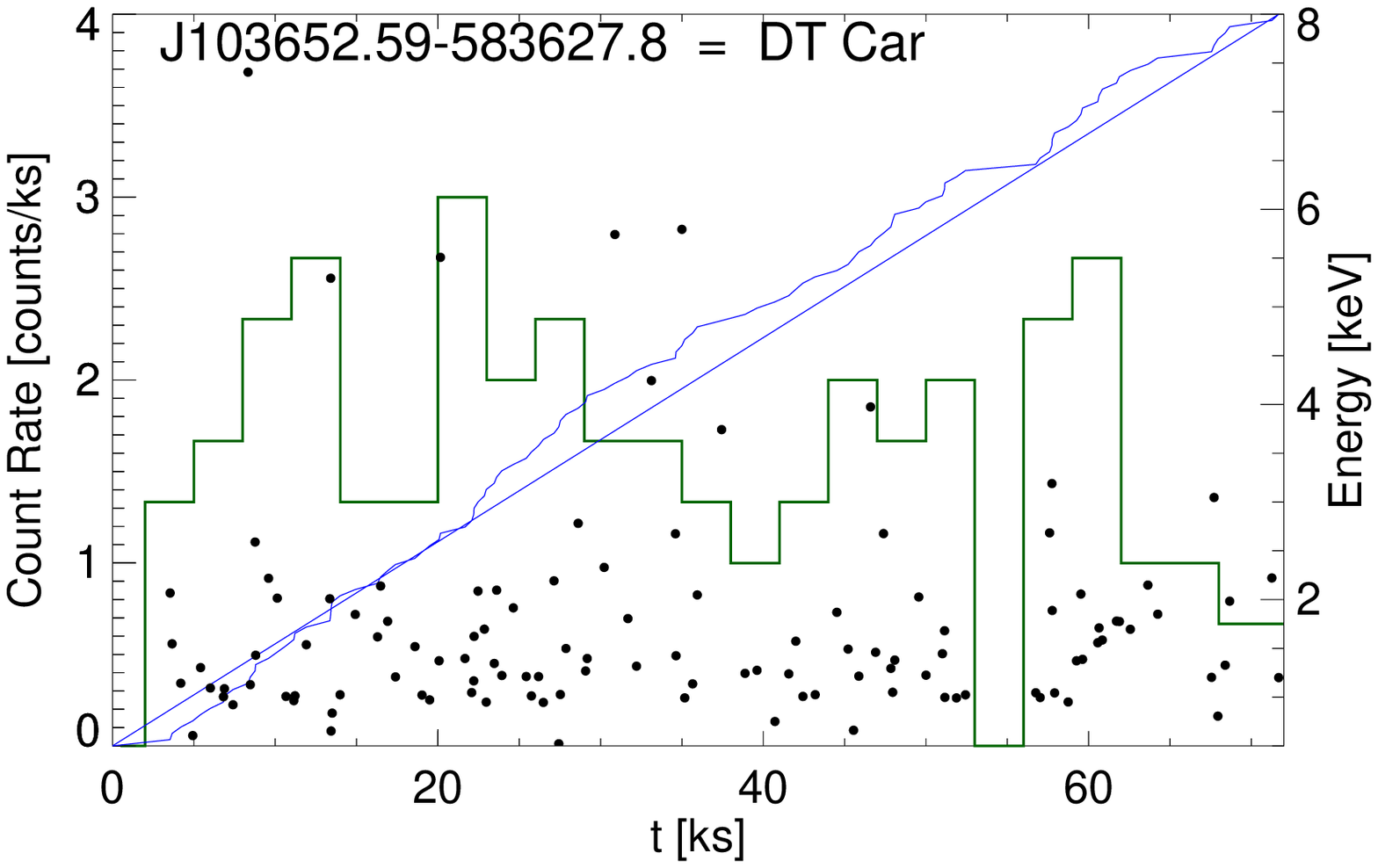}}
   \caption{\textit{Chandra} X-ray lightcurves of the two Algol stars
  KU~Car (J103745.21$-$584011.1) and DT~Car (J103652.59$-$583627.8).
  }
              \label{algols.fig}%
    \end{figure}

These two objects are detected as X-ray sources and here we provide a
brief discussion of their X-ray properties.

\paragraph{KU Car} is identified with the X-ray source J103745.21$-$584011.1.
The 27 X-ray source counts are not enough for spectral fitting,
but we note that the X-ray luminosity estimated with 
XPHOT is $\log \left( L_{\rm X}\,[{\rm erg/sec}] \right) \approx 31.0 $, if we assume a distance of 2.3~kpc.
The X-ray lightcurve of this source shows clear evidence
of a substantial increase in the count rate during the
$\approx 19$ hours of our observation.

It is interesting to compare the X-ray lightcurve to the
optical lightcurve of this system, for which \citet{Connell56}
determined a period of $5.921121 \pm 0.000007$ days.
According to the lightcurve elements derived in that study
(where the phase $\phi = 0$  corresponds to the minimum I),
the phase at the start of our \textit{Chandra} observation is $\phi = 0.54$,
and the phase at the end of our \textit{Chandra} observation is $\phi = 0.68$.
The minimum II (phase $\phi = 0.5$) happened 5.88~hours before start of the 
\textit{Chandra} observation.
According to the optical lightcurve shown in Fig.~5 of \citet{Connell56},
the visual brightness of the system increases within that period 
(i.e.,~it includes the final part of the occultation). 
The rise of the X-ray count rate during our \textit{Chandra}
observation is qualitatively
consistent with this and suggests that the low count rate
during the first hours of the observation are due to a
(partial) X-ray eclipse in the system.

\paragraph{DT Car,} the second known
Algol-type eclipsing binary in Gum~31, has a period of 4.2866 days
and a  spectral type of A2:+[G6Iv] as listed in SIMBAD.
This object is located just at the rim of the Gum~31 \ion{H}{ii} region.
It is detected as X-ray source J103652.59$-$583627.8 in our \textit{Chandra} data
with 117 source counts.
The X-ray luminosity derived from our spectral fit is 
 $\log \left( L_{\rm X}\,[{\rm erg/sec}] \right) = 31.06 \pm 0.16 $
(assuming a distance of 2.3~kpc).

According to the lightcurve elements listed in the VizieR catalog,
the range of phases covered by our \textit{Chandra} observation is from
$\phi = 0.45$ to $\phi = 0.65$. This means that the minimum II ($\phi = 0.5$) 
should have occurred 17.2~ks after the start of the observation.
However, no indication of variability is seen in the X-ray lightcurve at this time.
This may be related to uncertainties in the optical lightcurve elements;
no error bars are given for the period or the epoch of $\phi = 0$ in the
VizieR catalog, but we note that changing the last given digit of the period by 1 is enough
to shift the epoch of phase $\phi = 0.5$ outside the range of times covered
by our \textit{Chandra} observation.


\section{Properties of the X-ray selected young stellar population \label{pop.sect}}

\subsection{Infrared excesses of the X-ray sources}

   \begin{figure*}
   \centering
 \parbox{8.5cm}{\includegraphics[width=8.5cm]{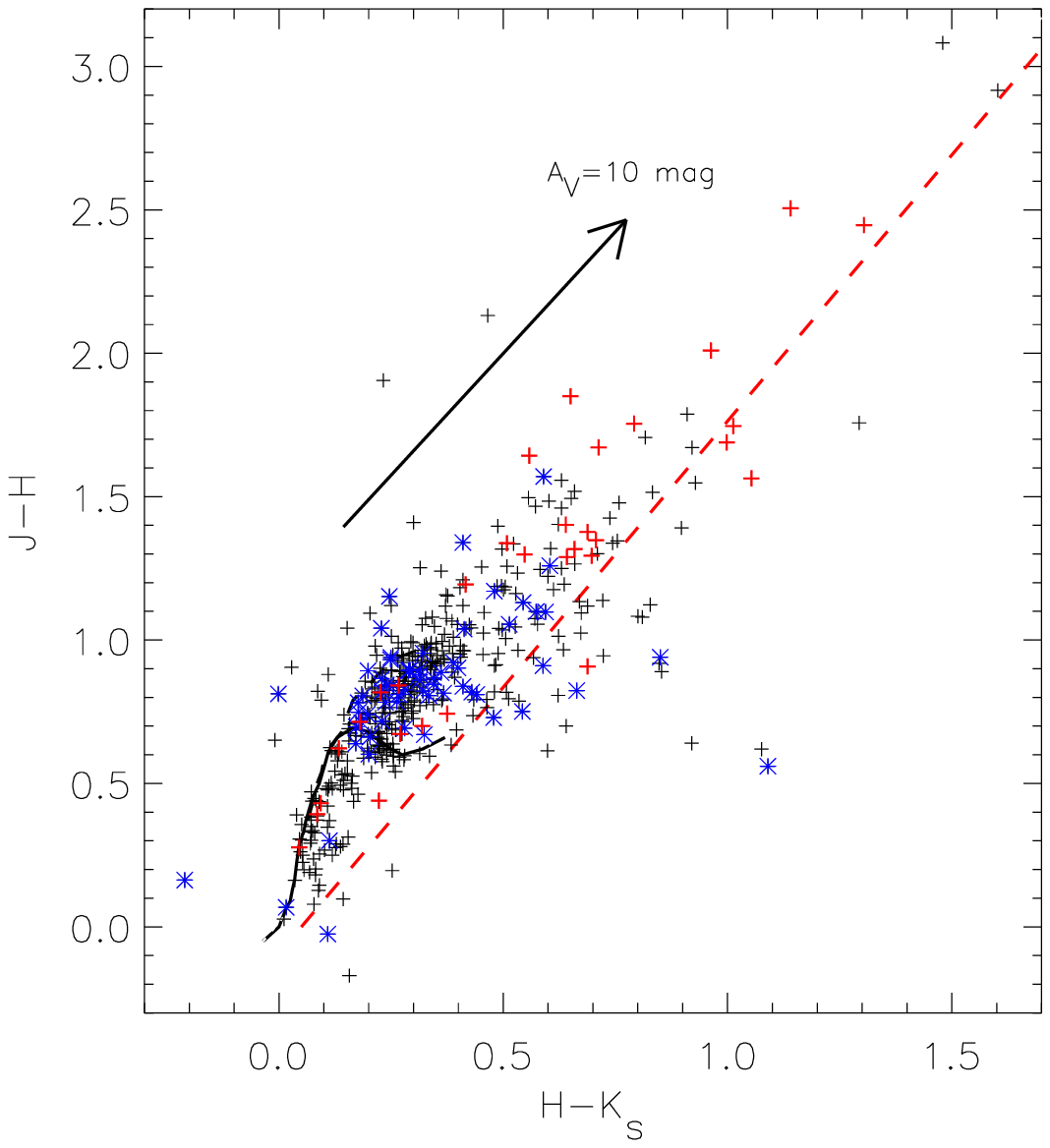}}\hspace{4mm}
 \parbox{8.5cm}{\includegraphics[width=8.5cm,height=9.2cm]{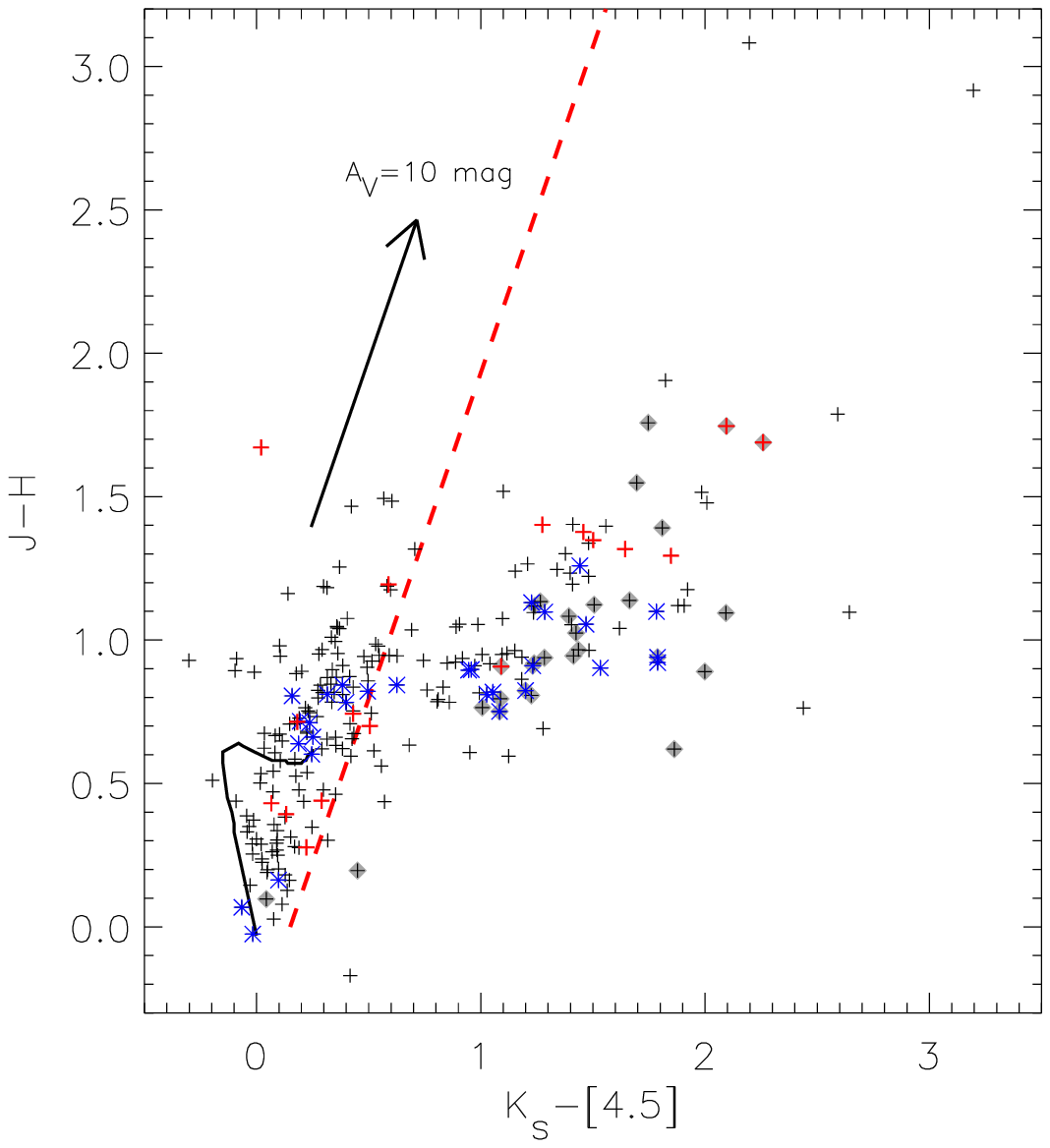}}
   \caption{Left: Near-infrared $J-H$ versus $H-K$ Color-Color Diagram of all X-ray selected sources
(crosses). The X-ray sources in the NGC~3324 cluster region are marked by blue asterisks, whereas
the sources in the G286.38--0.26 cluster region are marked by red crosses.
The solid line shows the main sequence,
the arrow shows a $A_V = 10$ mag reddening vector with slope 1.85,
and the red dashed line marks the separation between the photospheric reddening
band and the infrared excess region.
Objects are classified as near-infrared excess sources if
they lie at least 0.05~mag to the
right of the reddening band and above $J-H=0$.
\newline
Right: Near- to mid-infrared $J-H$ versus $K- [4.5]$ Color-Color Diagram of all X-ray selected sources.
The symbols have the same meaning as in the left plot.
Sources with NIR excesses are additionally
marked by filled grey diamonds.
  }
              \label{ccds.fig}%
    \end{figure*}

A quantification of the fraction of stars that are still surrounded by
circumstellar material (disks) can provide important information about
the evolutionary state of a region. A good way to do this is to
look for excesses in near- or mid-infrared color-color diagrams.

In Fig.~\ref{ccds.fig} we show a near-infrared $J-H$ vs.~$H-K_s$ 
color-color diagram for the X-ray selected sources.
Objects in this diagram are classified as NIR excess sources if
they lie at least 0.05~mag to the
right  of the reddening line, which is
based on the intrinsic colors of dwarfs \citep{BB88}
and an extinction vector with slope 1.85.
 In this way,
26 of the 472 X-ray sources with complete $J$, $H$, and $K_s$ photometry
are classified as NIR excess sources, i.e.,~we find a NIR excess fraction of
5.5\% in the X-ray selected population.

It is well known that
NIR emission only traces the hottest dust, close to the inner edge of the
disk. Therefore, 
the NIR excesses diminish rather quickly during the first few million years
of disk evolution.
Excesses in the mid-infrared range, which are caused by somewhat cooler disk material,
are known to persist considerably longer.
We have therefore also
constructed a color-color-diagram including
the mid-infrared IRAC2 $4.5\,\mu$m band.
As expected, the near/mid-infrared $J-H$ vs.~$K_s - [4.5]$
color-color diagram yields a considerably higher excess fraction
of 40.9\% (101 of the 247 objects with
complete photometry in the  $J$, $H$, $K_s$, and $[4.5]$ bands).

The NIR color-color diagram also allows us to characterize the extinction of the 
X-ray selected stars. Most objects show extinctions of about 2~mag, while
roughly 10\% of the sample have values of $A_V \ga 5$~mag, with extreme
values around $A_V \sim 20$~mag.

\subsection{Color-Magnitude Diagram of the X-ray sources}

   \begin{figure*}
   \centering
 \parbox{9cm}{\includegraphics[width=9cm]{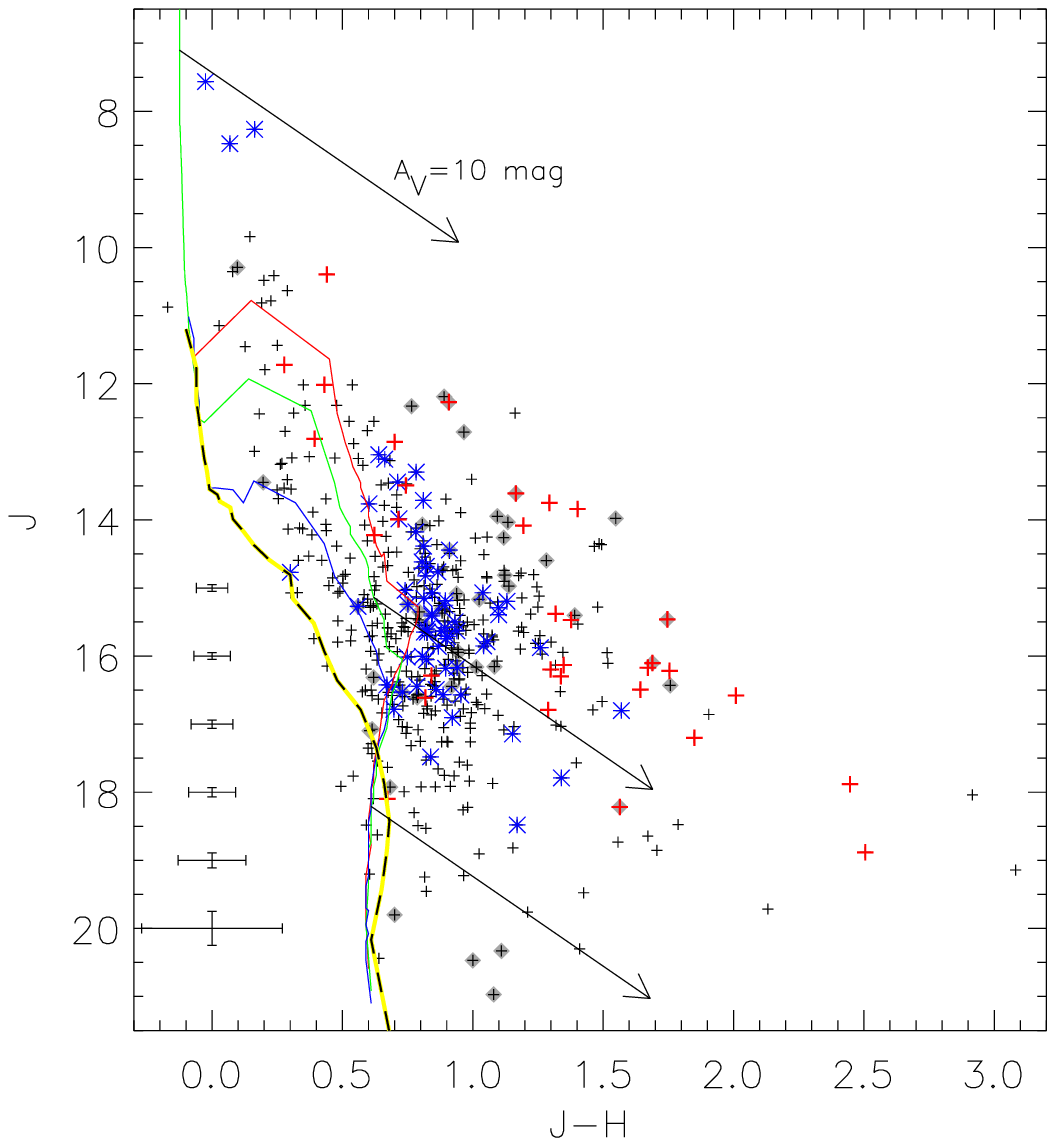}}
 \parbox{9cm}{\includegraphics[width=9cm]{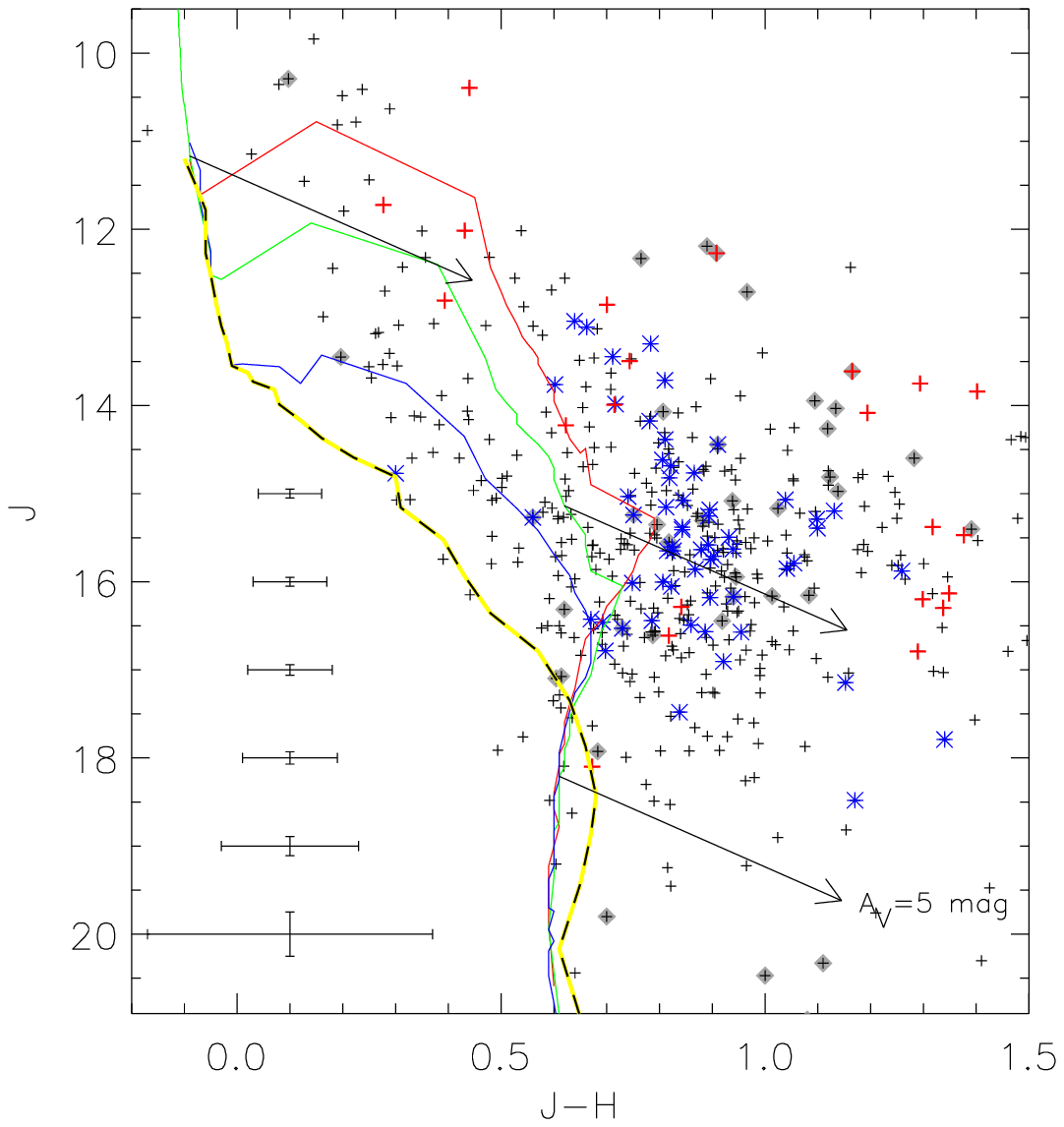}}
   \caption{Left: Near-infrared Color-Magnitude Diagram of the X-ray selected sources (crosses).
The X-ray sources in the NGC~3324 cluster region are marked by blue asterisks, whereas
the sources in the G286.38--0.26 cluster region are marked by red crosses.
Sources with NIR excesses are additionally
marked by filled grey diamonds.
The solid lines show isochrones for ages of
1~Myr (red), 3~Myr (green), and 10~Myr (blue)
composed from the models of \citet{Baraffe98} for the mass range
0.02 to 0.5~M$_\odot$, \citet{Siess00} for the
mass range 0.5 to 7~M$_\odot$, and \citet{Lejeune01} (model
 iso-c020-0650) for the mass range 7 to 70 ~M$_\odot$;
the dashed line shows the ZAMS from \citet{Siess00}.
The arrows indicate reddening vectors for $A_V = 10$~mag starting at the
location of 3~Myr old stars with masses of $35\,M_\odot$, $1\,M_\odot$, and
$0.1\,M_\odot$.
The row of error-crosses in the lower left part of the plot shows the
typical magnitude-dependent photometric uncertainties.
\newline
Right: Zoom into the central region of the CMD. 
The arrows indicate reddening vectors for $A_V = 5$~mag starting at the
location of 3~Myr old stars with masses of $7\,M_\odot$, $1\,M_\odot$, and
$0.1\,M_\odot$. }
              \label{cmds.fig}%
    \end{figure*}

Color-Magnitude Diagrams (CMDs) can be used to derive information about stellar ages and
masses. However, several factors, such as differential extinction,
infrared excesses, unresolved binary companions, and photometric variability,
can change the position of individual stars in the diagram
and thus strongly adulterate the derived stellar age and mass estimates.
The corresponding fundamental limitations of
age and mass determinations based on the
analysis of CMDs are discussed in \citet{HAWKI-survey} and 
also (in more detail) in \citet{Preibisch_ages}.
Nevertheless, it is possible to estimate the
{\em typical ages} for the different stellar populations in Gum~31.

The near-infrared CMD of the X-ray selected objects in Gum~31 is shown in 
Fig.~\ref{cmds.fig}.
The bulk of the X-ray stars have masses in the range $\sim 0.5\,M_\odot$ to
$\sim 3\,M_\odot$. Below  $\sim 0.5\,M_\odot$, the number drops strongly,
in good agreement with the expectation from the X-ray detection limit.
Just a few
objects are seen at locations corresponding to masses around $\sim 0.1\,M_\odot$.

The positions of the three O-type stars suggest moderate extinctions of
$A_V \sim 2$~mag for these objects; this agrees very well with the
column densities derived from the fits to the X-ray spectra of
HD~92\,206~A and B ($N_{\rm H} \approx 4 \times 10^{21}\,{\rm cm}^{-2}$, 
corresponding to $A_V \approx 2$~mag for both objects).

\subsection{CCD and CMD results for the cluster NGC~3324}

The CCD suggests  
extinctions of typically $A_V \sim 2$~mag for the X-ray selected objects in this cluster.
The NIR excess fraction is 9.5\%
(6 out of 63 objects with  complete photometry in the  $J$, $H$, and $K_s$ bands),
while the MIR  excess fraction is 51.7\%
(15 out of 29 objects with  complete $J$, $H$, $K_s$, and $[4.5]$ photometry).

In the CMD, almost all stars in NGC~3324 
are located to the right of the 1~Myr isochrone.
There are only two exceptions: one is source J103721.95$-$583708.0, that appears
close to the 10~Myr isochrone, the other one is J103714.63$-$583705.5 that appears at the ZAMS.
We note that J103721.95$-$583708.0 shows a clear NIR excess, and thus its
position in the CMD can not be used to infer an age.
This leaves J103714.63$-$583705.5 as the only object in the cluster area
that appears clearly offset (in the CMD) from the rest of the cluster
population; this suggests that this object may not be a member of this cluster, but
is only seen there due to projection effects. 
For example, it could be a reddened background O-type star.

Taking the typical reddening of $A_V \sim 2$~mag into account, these locations 
of the stars in the CMD suggest an age of $\la 2$~Myr. 
The fact that the cluster region is free from any 
dense clouds, i.e.,~that 
the cluster stars must have already completely dispersed
their natal cloud, implies that the age should be not much less that
$\ga 1$~Myr.

The NIR excess fraction of 9.5\% and the MIR excess rate of 51.5\% 
are consistent with an age of $\sim 2$~Myr \citep[see, e.g.,][]{Fedele10}.

Therefore, our final age estimate for NGC~3324 is  $\approx 1-2$~Myr.
We note that this value agrees  well to the 
expansion age of Gum~31 \ion{H}{ii} region, which has been estimated
to be about 1.5 Myr \citep{Ohlendorf13}.

\subsection{Results for the cluster G286.38--0.26}

The CCD shows a roughly bimodal distribution of extinctions: 
several stars have rather low extinction values ($A_V \la 1$~mag), whereas
several others show substantial extinction of $A_V \ga 5$~mag.
The NIR excess fraction is 16.1\%
(5 out of 31 objects with  complete photometry in the  $J$, $H$, and $K_s$ bands),
while the MIR  excess fraction is 52.9\%
(9 out of 17 objects with  complete $J$, $H$, $K_s$, and $[4.5]$ photometry).

All but three of the stars in the cluster area are located 
to the right of the 1~Myr isochrone in the CMD.
These three apparently older stars are in the upper part of the CMD, which indicates ages of
about $2-3$~Myr; these objects belong to the group of bright 
optically visible stars in the region.

All other X-ray sources in this region appear to have ages of
$\la 1$~Myr and most of them show substantial amounts of extinction.
Since there are large amounts of dense cloud material left, in which
several deeply embedded young stellar objects are found,
star formation is clearly going on in this cluster.
Our final final age range estimate for G286.38--0.26 is therefore
$\la 1 - 2$~Myr.
The NIR excess fraction of 16\%  and MIR excess fraction of 53\% 
is consistent with this.

\subsection{The distributed population}

We finally discuss the
population of X-ray selected stars in the \ion{H}{II} region that 
is not located in the clusters NGC~3324 and G286.38--0.26.

The CMD of the distributed stars  
is clearly different from the CMDs of the clusters NGC~3324 and G286.38--0.26
because it shows a broader spread of colors.
The considerable number of objects to the left of the 3~Myr isochrone
and the presence of several objects near the 10~Myr isochrone
suggest a rather broad age distribution, with ages of up to $\sim 10$~Myr for
this population.
The excess rates of only 4\% (NIR) and 38\% (MIR)
also suggest older ages for at least some part of this population.

\section{Global properties of the young stellar population in Gum~31}

\subsection{The size of the X-ray selected population}

   \begin{figure}
   \centering
 \includegraphics[width=8cm]{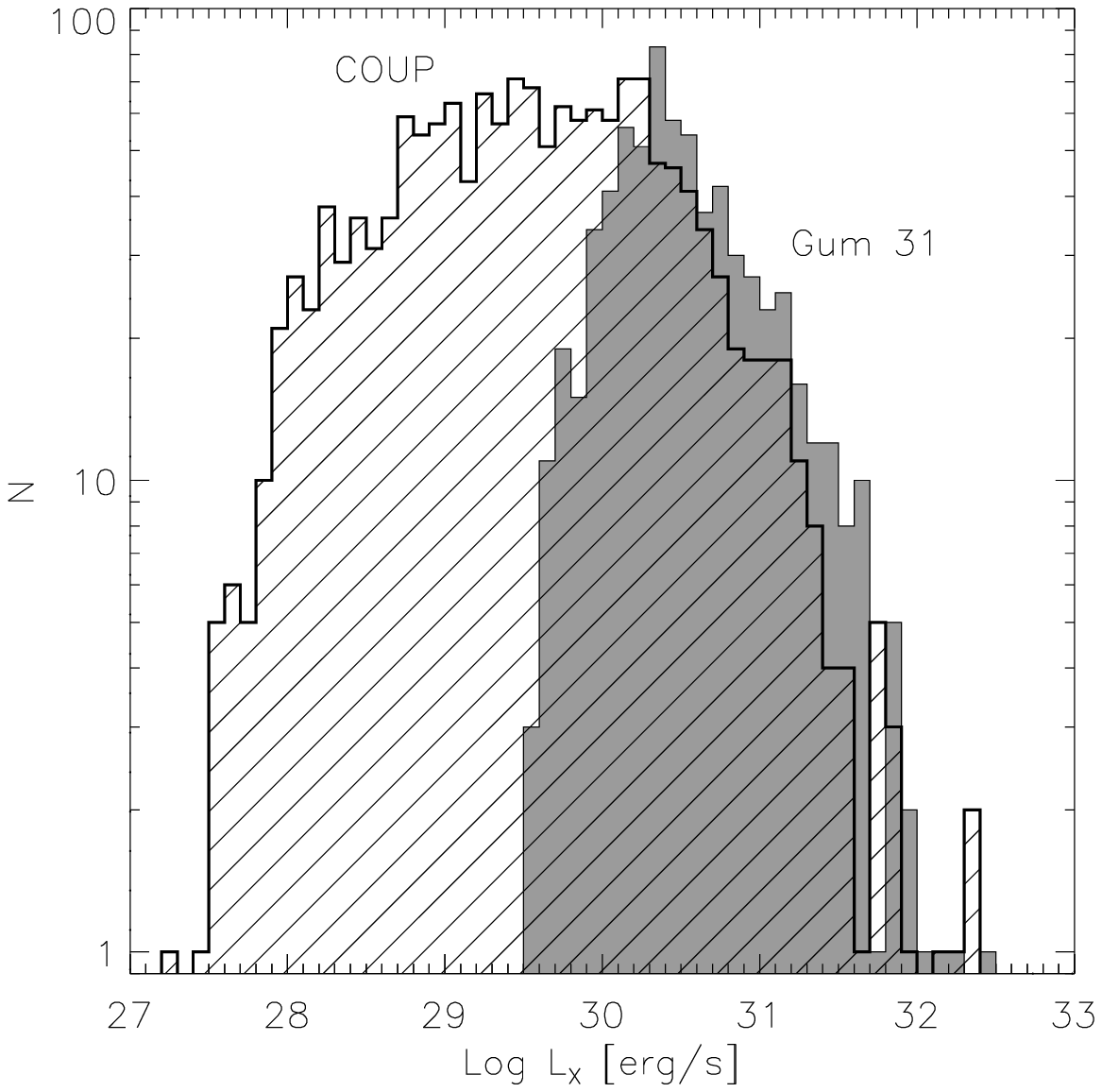}
 \includegraphics[width=8cm]{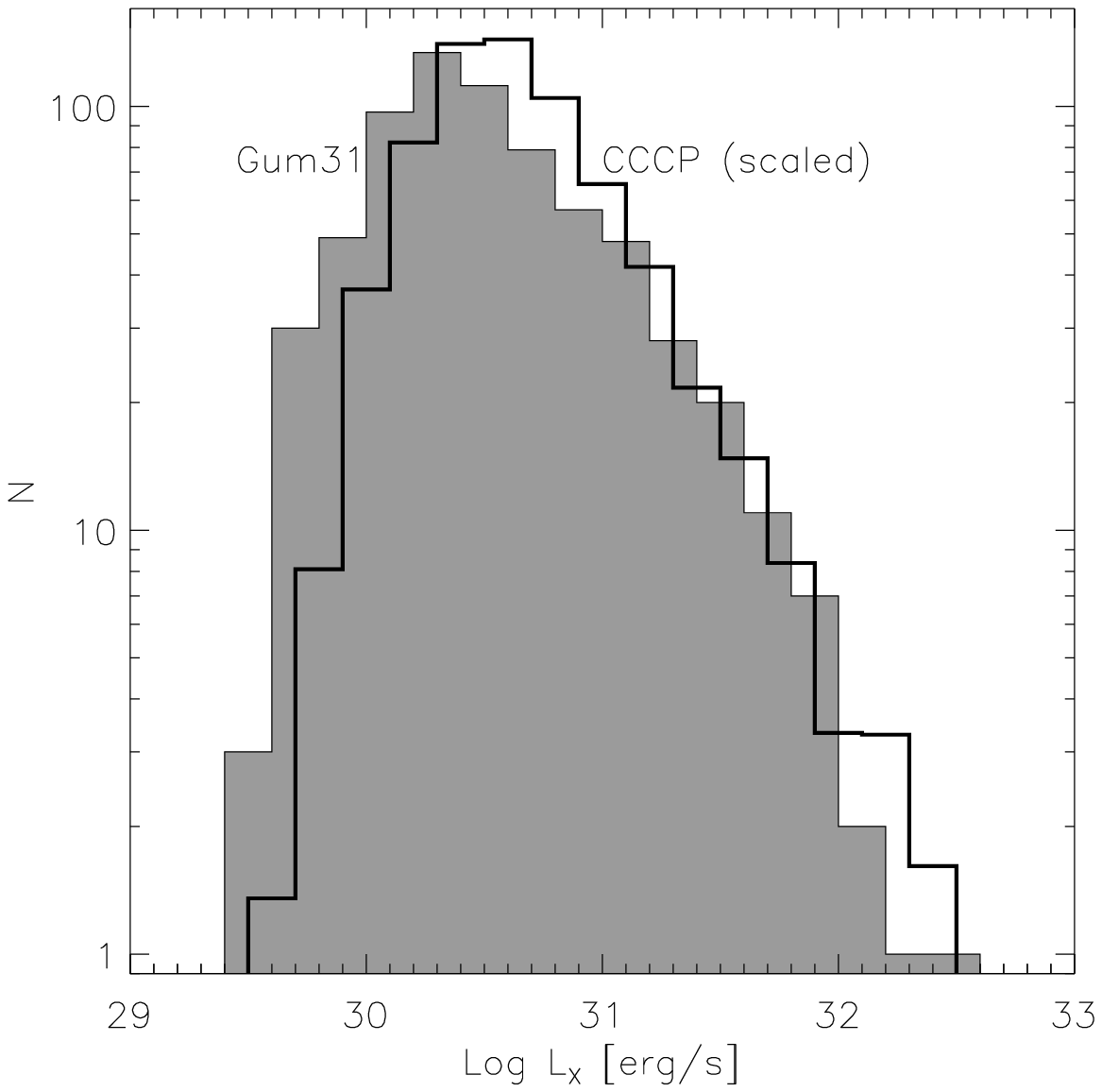}
\caption{Comparison of the X-ray luminosity function
 of the \textit{Chandra} sources in our Gum~31 observation
 (grey-shaded histogram) to
({\bf top:}) the XLF of the
Orion nebula Cluster (dashed histogram) 
\citep[from the COUP data;][]{get05},
and to ({\bf bottom:}) the XLF for the CCCP sources as shown
in Fig.~3 of \citet{CCCP-Clusters}, scaled to account
for the different sample sizes.
  }
              \label{xlf.fig}%
    \end{figure}

In the study of \citet{CCCP-Clusters}, the observed
distribution of X-ray luminosities (the so-called
``X-ray luminosity function'', \textit{XLF}) of the sources in 
different parts of the Carina nebula was used to
derive estimates of the underlying total stellar populations
by comparison to the XLF of the Orion nebula Cluster (ONC),
which was particularly well studied in X-rays in the 
context of the \textit{Chandra Orion Ultradeep Project} (COUP)
\citep[see][]{get05}.

We use a similar strategy here to obtain information about the
total population of young stars in the Gum~31 area.
In Fig.~\ref{xlf.fig} (upper panel) we compare the distribution of 
 X-ray luminosity estimates
(see Sect.~2.4) for our Gum~31 sample to the 
COUP data.
Both distributions show a qualitatively similar increase in the
source numbers when going from the highest towards lower
X-ray luminosities. As a consequence of the larger target distance and
the shorter exposure time, the Gum~31 distribution turns over
and drops down at a much higher luminosities 
($\log \left( L_{\rm X}\,[{\rm erg/sec}] \right) \approx 30.3$)
than does the COUP data.
Above this limit, however, the ratio of the number of X-ray sources
should be approximately equal the the ratio of the respective
population sizes.

For the Gum~31 sample, we count 445 X-ray sources in the
luminosity range $\log \left( L_{\rm X}\,[{\rm erg/sec}] \right) = [30.3 \dots 32.0]$,
whereas the COUP sample has 305 objects in this range.
Therefore, the
number of X-ray emitting stars in Gum~31 is about 1.46 times higher
than in the ONC. If we assume the ONC population to consist of
2800 stars \citep[see][]{CCCP-Clusters}, the predicted population in our Gum~31 field is 4085.

In the lower panel of Fig.~\ref{xlf.fig} we compare the Gum~31 XLF
to the XLF for the CCCP sources as shown
in Fig.~3 of \citet{CCCP-Clusters} (scaled to account for 
the difference in sample size).
In the luminosity range $\log \left( L_{\rm X}\,[{\rm erg/sec}] \right) \approx [31 \dots 32]$,
the distributions agree well.
The lower number of very bright ($\log \left( L_{\rm X}\,[{\rm erg/sec}] \right) > 32$)
sources in the Gum~31 XLF seems to be related to the smaller number of X-ray bright O-type stars.

\subsection{Clustered versus distributed population}

As mentioned above,
the spatial distribution of the X-ray sources shows two prominent
clusters, as well as a homogeneously distributed population. 
The sizes of the clustered
and the distributed population of young stars can be 
estimated as follows.
About 150 of the $\approx 500$ young stars are located in the clusters
NGC~3324 or G286.38--0.26, or in one of the smaller clusterings discussed in 
Sect.~2.5. The remaining $\approx 350$ sources constitute the
distributed population in the Gum~31 region.
Their spatial distribution  shows a rather uniform
surface density and
there is no obvious large-scale density gradient towards the edges of the
observed area (above the expected level resulting from the decreasing
instrumental sensitivity with increasing offaxis-angle).
This suggests that this distributed population probably extends beyond
the field of our \textit{Chandra} observation.
The clustered and the distributed population show also differences
with respect to their stellar ages: while the clusters seem to be 
quite young ($\la 2$~Myr), 
about 20\% of the objects in the distributed population have ages
of more than $\approx 3$~Myr, up to about 10~Myr.

A similar bimodal spatial configuration was found for the 
X-ray sources in the CCCP field (i.e.,~the more central parts
of the Carina nebula): about half of the probable Carina members in the 
CCCP sample are members in one of several clusters, while the other half
constitutes a so-called
``widely-distributed population'', which is spread out through the
entire CCCP area \citep{CCCP-Clusters}.
In \citet{CCCP-HAWKI} and \citet{HAWKI-survey} we found that 
this widely-distributed population
of young stars shows a relatively wide range of stellar ages, up to
$\sim 10$~Myr.

These similarities raise the question, how the distributed population
we find in the Gum~31 region is related to the 
widely-distributed population in the CCCP field. 
Using the numbers given in \citet{CCCP-Clusters}, we find that the surface 
density\footnote{Table~1 of \citet{CCCP-Clusters} lists 
5271 probable Carina members in the CCCP sample  that
are located outside the large-scale stellar enhancements A, B, C
(which contain the different identified clusters). 
The area of this
``Region D'' is $\approx 1.27$  square-degrees.}
of the widely-distributed population in the CCCP field
is $\approx 1.15$ stars per square-arcminute.
This predicts $\approx 330$ stars per single ACIS-I field, and this
number is in good agreement with the size of the
distributed population we find in our Gum~31 region.

It is thus feasible that the distributed population of young stars
in Gum~31 is an extension of the widely-distributed population in the
inner parts of the Carina nebula.
Further implications of this assumption will be discussed in Sect.~9.

\section{Diffuse X-ray emission}

   \begin{figure}
   \centering
\includegraphics[width=8.5cm]{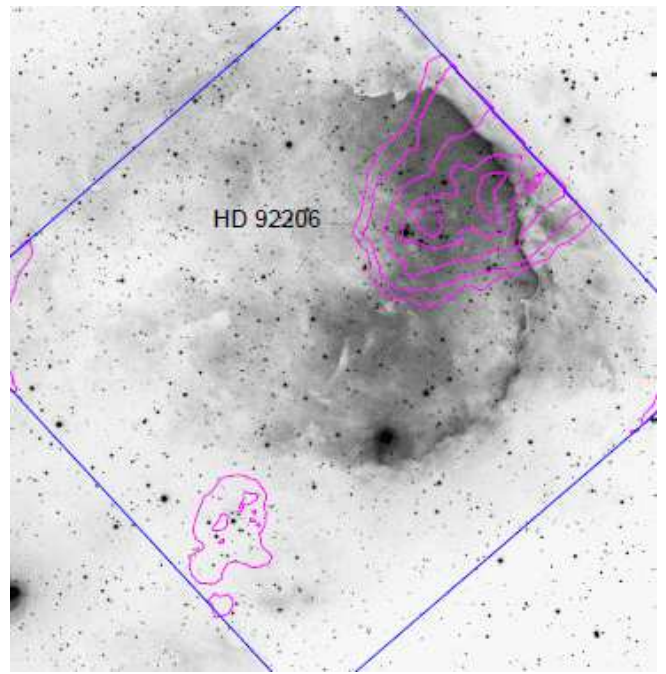}
   \caption{Negative representation of the WFI $R$-band image 
(see {\tt www.eso.org/public/images/eso1207a/}; image credit: ESO)
with superposed
contours of the smoothed apparent surface brightness of the diffuse X-ray emission 
in the $0.5-7$~keV band (point-sources have been excised).
The contour levels are drawn at levels of
 $1.9, 2.1, 2.6, 3.4,\,{\rm and}\,4.0 \times 10^{-9}\,{\rm photons}\,{\rm s}^{-1}\,
{\rm cm}^{-2}\,{\rm arcsec}^{-2}$.
The blue lines mark the borders of the $17' \times 17'$
\textit{Chandra}/ACIS-I field-of-view. The position of the
O-star multiple system HD~92\,206 is marked.}
              \label{fig:diffuse}%
    \end{figure}

The strong  stellar winds of massive stars can fill the area
around a star or a stellar cluster with a hot plasma that can be 
traced by diffuse soft X-ray emission.
Such a diffuse X-ray component has been discovered in several
high-mass star forming regions \citep[see, e.g.,][]{Guedel08,CCCP-diffuse2}.
In the central parts of the CNC, strong diffuse  X-ray emission
has been detected in the CCCP survey and is described in detail in
\citet{CCCP-diffuse1,CCCP-diffuse2}.

We have performed a search for diffuse X-ray emission in the Gum~31 region
in a similar way. We excised all point sources, then created X-ray images in 
several different energy bands and smoothed them with an adaptive-kernel smoothing code
\citep{Broos10}.
The resulting smoothed diffuse X-ray image of Gum~31
for the $0.5-7$~keV band is shown in Fig.~\ref{fig:diffuse}.
One can clearly see significant levels of diffuse X-ray emission
at several locations in Gum~31.
The brightest patches of diffuse X-ray emission are found in a region just northwest of the
stellar cluster NGC~3324. This suggests that the winds of the three O-type stars
in NGC~3324 produce this hot plasma. The emission shows a very inhomogeneous spatial distribution,
strongly concentrated to the northwest and much weaker in the other directions
(as measured from the position of the cluster). 
The plasma flowing towards the northwest seems to be
relatively well confined by the clouds surrounding the \ion{H}{ii} bubble,
what leads to relatively high plasma densities and correspondingly strong
X-ray emission at this location.

A second peak of diffuse X-ray emission is seen in the area of the
partly embedded cluster G286.38--0.26.
The origin of this plasma is not as easily explained, since no
O-type stars are known to be present here.

A more detailed analysis of the diffuse X-ray emission in Gum~31
will be performed in connection with 
our new submm mapping data in a forthcoming publication.
A detailed X-ray spectral analysis of the diffuse emission will allow us to
estimate plasma temperatures and densities and provide more quantitative
information.

\section{Conclusions and Summary}

Our \textit{Chandra} observation of the Gum~31 region lead to the
detection of 679 X-ray point sources, about 500 of which are young stars.
This allowed us to identify, for the first time,
a large sample of the young stellar population
in the Gum~31 area, which is complete down to $\sim 1\,M_\odot$.
Extrapolation of the X-ray luminosity function down to $0.1\,M_\odot$
suggests a total population of about 4000 young stars in the observed
area.
This shows that the still poorly explored northern parts of the
CNC contain a very substantial fraction of the total
young stellar population in the complex.

About 30\% of the $\sim 500$ X-ray detected young stars
in the Gum~31 area are concentrated in the two very young ($\la 2$~Myr) 
clusters NGC~3324 and G286.38--0.26, but 
the large majority (70\%)
is roughly uniformly distributed throughout the observed 
field-of-view.
The analysis of the near-infrared color-magnitude diagram suggests
that these distributed stars have a range of ages up to 
about 10~Myr.

Interestingly, the surface density of the distributed population in Gum~31 is similar
to that of the ``widely distributed population'' that was found in the 
inner parts of the Carina nebula. 
This may suggest that the distributed population of young stars 
in the CNC extends over (at least) 100 pc.
Such a large-scale distributed population would then contain
several times more stars than the total number of members in all the
known clusters (like Tr14, 15, 16, and as NGC~3324).
These clusters clearly dominate the visual impression of the region,
e.g.,~by creating the optically prominent \ion{H}{ii} regions.
However, our results here suggest that 
most of the stellar mass is actually in the much more inconspicuous
large-scale distributed stellar population.
The Carina nebula complex is thus a stellar association, in which
the known clusters represent just a few compact subgroups.
This is probably the consequence of the stochastic dynamical evolution
 \citep[see, e.g.,][]{Parker14}
in combination with the extended history of star formation in the complex.

Further observations are required to quantify and finally evaluate this scenario.
A spatially complete and deep enough X-ray survey would be 
the ideal tool, but this is not really feasible.
However, recent wide-field infrared surveys will soon provide
more clues.

\begin{acknowledgements}

We thank the referee for a very careful and constructive report that helped to
improve this paper.

This work was supported by funding from Deutsche Forschungsgemeinschaft under
DFG project number PR~569/9-1. Additional support came
from funds from the Munich Cluster of Excellence ``Origin and Structure of the
Universe''.

The scientific results reported in this article are based on observations 
made by the \textit{Chandra} X-ray Observatory.
Support for this work was provided by the National Aeronautics and Space Administration 
through Chandra Award Number GO2-13010X issued by the \textit{Chandra}
 X-ray Observatory Center, 
which is operated by the Smithsonian Astrophysical Observatory for and on behalf of the 
National Aeronautics Space Administration under contract NAS8-03060.
This research has made use of software provided by the \textit{Chandra}
 X-ray Center (CXC) in the application packages CIAO, ChIPS, and Sherpa.

This work uses data from observations made with ESO Telescopes at the La Silla Paranal 
Observatory under program ID 088.C-0117(A).

This work is based in part on observations made with the \textit{Spitzer} Space Telescope, 
which is operated by the Jet Propulsion Laboratory, California Institute of Technology under a contract with NASA.

This research has made use of the SIMBAD database,
operated at CDS, Strasbourg, France.
\end{acknowledgements}


\begin{thebibliography}{}

\bibitem[Anders \& Grevesse(1989)]{ag89} Anders, E., \& Grevesse, N.\ 1989, \gca, 53, 197

\bibitem[Baraffe et~al.(1998)]{Baraffe98} Baraffe I., Chabrier G., Allard F., 
Hauschildt P.H.\ 1998, A\&A, 337, 403

\bibitem[Bessel \& Brett(1988)]{BB88} Bessel, M.S., Brett, J.M.\ 1988, PASP, 100, 1134

\bibitem[Brice{\~n}o et al.(2007)]{Briceno07} Brice{\~n}o C., Preibisch, Th.,
Sherry, W., Mamajek, E., Mathieu, R., Walter, F., Zinnecker, H.\ 2007,
 in:
Protostars \& Planets V, eds. B. Reipurth, D. Jewitt, \& K. Keil,
University of Arizona Press, Tucson, p.~345 

\bibitem[Brooks et~al.(2003)]{Brooks03} Brooks, K.J., Cox, P., Schneider, N., Storey, J.~W.~V., 
Poglitsch, A., Geis, N., \& Bronfman, L.\ 2003, A\&A, 412, 751

\bibitem[Brooks et~al.(2005)]{Brooks05} Brooks, K.J., Garay, 
G., Nielbock, M., Smith, N., \& Cox, P.\ 2005, ApJ, 634, 436

\bibitem[Broos et al.(2007)]{bro07} Broos, P.~S., Feigelson, 
E.~D., Townsley, L.~K., et al.\ 2007, \apjs, 169, 353 

\bibitem[Broos et al.(2010)]{Broos10} Broos, P.~S., Townsley, 
L.~K., Feigelson, E.~D., et al.\ 2010, \apj, 714, 1582 

\bibitem[Broos et al.(2011a)]{CCCP-catalog} Broos, P.~S., Townsley, L.~K., Feigelson, E.~D., et al.\ 2011a, \apjs, 194, 2 

\bibitem[Broos et al.(2011b)]{CCCP-classification} Broos, P.~S., Getman, K.V., Povich, M.S., et al.\ 2011b, \apjs, 194, 4

\bibitem[Broos et al.(2012)]{AE2012} Broos, P., Townsley, L., 
Getman, K., \& Bauer, F.\ 2012, Astrophysics Source Code Library, 3001 

\bibitem[Campillay et al.(2007)]{Campillay07} Campillay, A., Arias, 
J., Barba, R., et al.\ 2007, VI Reunion Anual Sociedad Chilena de 
Astronomia (SOCHIAS), 63 

\bibitem[Cappa et al.(2008)]{Cappa08} Cappa, C., Niemela, V.~S., Amor{\'{\i}}n, R., \& Vasquez, J.\ 2008, \aap, 477, 173 

\bibitem[Carraro et al.(2001)]{Carraro01} Carraro, G., Patat, F., \& Baumgardt, H.\ 2001, \aap, 371, 107 

\bibitem[Casertano \& Hut(1985)]{CH85} Casertano, S., \& Hut, P.\ 1985, \apj, 298, 80 

\bibitem[O'Connell(1956)]{Connell56} O'Connell, D.~J.~K.\ 1956, 
Ricerche Astronomiche, 3, 313 

\bibitem[Dalton et al.(2006)]{Dalton06} Dalton, G.~B., Caldwell, 
M., Ward, A.~K., et al.\ 2006, \procspie, 6269,  

\bibitem[Dias et al.(2002)]{Dias02} Dias, W.~S., Alessi, B.~S., Moitinho, A., 
\& L{\'e}pine, J.~R.~D.\ 2002, \aap, 389, 871 

\bibitem[Dutra et al.(2003)]{Dutra03} Dutra, C.~M., Bica, E., Soares, J., \& Barbuy, B.\ 2003, \aap, 400, 533 

\bibitem[Emerson et al.(2006)]{Emerson06} Emerson, J., McPherson, 
A., \& Sutherland, W.\ 2006, The Messenger, 126, 41 

\bibitem[Fedele et al.(2010)]{Fedele10} Fedele, D., van den Ancker, M.~E., Henning, T., 
Jayawardhana, R., \& Oliveira, J.~M.\ 2010, \aap, 510, A72 

\bibitem[Feigelson et al.(2007)]{Feigelson07} Feigelson, E.~D., 
Townsley, L.,
G\"udel, M., Stassun, K. \ 2007, Protostars \& Planets V,
(eds. B. Reipurth, D. Jewitt, and K. Keil, Univ. Arizona Press), p.~313


\bibitem[Feigelson et al.(2011)]{CCCP-Clusters} Feigelson, E.~D., Getman, K.~V., Townsley, L.~K., et al.\ 2011, \apjs, 194, 9 

\bibitem[Freeman et al.(2001)]{Freeman01} Freeman, P., Doe, S., 
\& Siemiginowska, A.\ 2001, \procspie, 4477, 76

\bibitem[Freeman et al.(2002)]{Freeman02} Freeman, P.~E., 
Kashyap, V., Rosner, R., \& Lamb, D.~Q.\ 2002, \apjs, 138, 185 

\bibitem[Gaczkowski et al.(2013)]{Gaczkowski13} Gaczkowski, B., Preibisch, T., Ratzka, T., et al.\ 2013, \aap, 549, A67 

\bibitem[Gagn{\'e} et al.(2011)]{Gagne11} Gagn{\'e}, M., Fehon,               
G., Savoy, M.~R., et al.\ 2011, \apjs, 194, 5

\bibitem[Garmire et al.(2003)]{Garmire03} Garmire, G.~P., Bautz, 
M.~W., Ford, P.~G., Nousek, J.~A., 
\& Ricker, G.~R., Jr.\ 2003, \procspie, 4851, 28 

\bibitem[Gehrels(1986)]{Gehrels86} Gehrels, N.\ 1986, \apj, 303, 336 

\bibitem[Getman et al.(2005)]{get05} Getman, K.~V., 
Feigelson, E.~D., Grosso, N., et al.\ 2005, \apjs, 160, 353 

\bibitem[Getman et al.(2010)]{get10} Getman, K.~V., 
Feigelson, E.~D., Broos, P.~S., Townsley, L.~K., 
\& Garmire, G.~P.\ 2010, \apj, 708, 1760 

\bibitem[Getman et al.(2011)]{Getman11} Getman, K.~V., Broos, 
P.~S., Feigelson, E.~D., et al.\ 2011, \apjs, 194, 3 

\bibitem[G{\"u}del et al.(2008)]{Guedel08} G{\"u}del, M., 
Briggs, K.~R., Montmerle, T., et al.\ 2008, Science, 319, 309 

\bibitem[G{\"u}del et al.(2007)]{Guedel07} G{\"u}del, M., Briggs, K.~R., 
Arzner, K., et al.\ 2007, \aap, 468, 353 

\bibitem[Gum(1955)]{Gum55} Gum, C.~S.\ 1955, \memras, 67, 155 

\bibitem[Gutermuth et al.(2009)]{Gutermuth09} Gutermuth, R.~A., 
Megeath, S.~T., Myers, P.~C., et al.\ 2009, \apjs, 184, 18 

\bibitem[Irwin et al.(2004)]{Irwin04} Irwin, M.~J., Lewis, J., 
Hodgkin, S., et al.\ 2004, \procspie, 5493, 411 

\bibitem[Kroupa(2002)]{Kroupa02} Kroupa, P.\ 2002, Science, 295, 82

\bibitem[Lejeune \& Schaerer(2001)]{Lejeune01} Lejeune T., 
Schaerer D.\ 2001, A\&A, 366, 538 

\bibitem[Lucy(1974)]{Lucy74} Lucy, L.~B.\ 1974, \aj, 79, 745 

\bibitem[Martins et al.(2005)]{Martins05} Martins, F., Schaerer, D., \& Hillier, D.~J.\ 2005, \aap, 436, 1049 

\bibitem[Mathys(1988)]{Mathys88} Mathys, G.\ 1988, \aaps, 76, 427 

\bibitem[Ohlendorf et al.(2012)]{Ohlendorf12} Ohlendorf, H., Preibisch, T., Gaczkowski, B., et al.\ 2012, \aap, 540, A81

\bibitem[Ohlendorf et al.(2013)]{Ohlendorf13} Ohlendorf, H., Preibisch, T., Gaczkowski, B., et al.\ 2013, \aap, 552, A14 

\bibitem[Ohlendorf et al.(2014)]{Ohlendorf14} Ohlendorf, H., Preibisch, T., Roccatagliata, V.,
Ratzka, T.\ 2014, \aj, submitted

\bibitem[Parker et al.(2014)]{Parker14} Parker, R.~J., Wright, 
N.~J., Goodwin, S.~P., \& Meyer, M.~R.\ 2014, \mnras, 438, 620 

\bibitem[Povich et al.(2011)]{Povich11} Povich, M.~S., Smith, N., Majewski, S.~R., et al.\ 2011, \apjs, 194, 14 

\bibitem[Preibisch(2012)]{Preibisch_ages} Preibisch, T.\ 2012, 
Research in Astronomy and Astrophysics, 12, 1 

\bibitem[Preibisch \& Feigelson(2005)]{PF05} Preibisch, T., \& Feigelson, E.~D.\ 2005, \apjs, 160, 390 

\bibitem[Preibisch et al.(2005)]{Preibisch_coup_orig} Preibisch, Th., 
Kim, Y.-C., Favata, F., et al.\ 2005a, \apjs, 160, 401

\bibitem[Preibisch et al.(2011a)]{CNC-Laboca} Preibisch, T., Schuller, F., Ohlendorf, H., et al.\ 2011a, \aap, 525, A92 

\bibitem[Preibisch et al.(2011b)]{CCCP-HAWKI} Preibisch, T., Hodgkin, S., Irwin, M., et al.\ 2011b, \apjs, 194, 10 

\bibitem[Preibisch et al.(2011c)]{HAWKI-survey} Preibisch, T., Ratzka, T., Kuderna, B., et al.\ 2011c, \aap, 530, A34 

\bibitem[Preibisch et al.(2012)]{Preibisch12} Preibisch, T., Roccatagliata, V., Gaczkowski, B., \& Ratzka, T.\ 2012, \aap, 541, A132 


\bibitem[Roccatagliata et al.(2013)]{Roccatagliata13} Roccatagliata, V., Preibisch, T., 
Ratzka, T., \& Gaczkowski, B.\ 2013, \aap, 554, A6

\bibitem[Siess et al.(2000)]{Siess00} Siess, L.,
Dufour, E., \& Forestini, M.\ 2000, \aap, 358, 593

\bibitem[Smith(2006)]{Smith06} Smith, N.\ 2006, \mnras, 367, 763 

\bibitem[Smith \& Brooks(2007)]{SB07} Smith, N., \& Brooks, K.~J.\ 2007, \mnras, 379, 1279 

\bibitem[Smith \& Brooks(2008)]{SB08} Smith, N. \& Brooks, 
K.J.\ 2008, in: Handbook of Star Forming Regions, 
Volume II: The Southern Sky, ASP Monograph Publications, Vol.~5.~Edited by Bo Reipurth, p.~138

\bibitem[Smith et~al.(2010a)]{Smith10a} Smith, N., Bally, J., \&
Walborn, N.R.\ 2010a, \mnras, 405, 1153 

\bibitem[Smith et~al.(2010b)]{Smith10b} Smith, N., Povich, M.S., 
Whitney, B.A., et al.\ 2010b, \mnras, 406, 952 

\bibitem[Stassun et al.(2006)]{Stassun06} Stassun, K.~G., van den 
Berg, M., Feigelson, E., \& Flaccomio, E.\ 2006, \apj, 649, 914 

\bibitem[Stelzer et al.(2005)]{Stelzer05} Stelzer, B., Flaccomio, 
E., Montmerle, T., et al.\ 2005, \apjs, 160, 557

\bibitem[Townsley et al.(2003)]{ton03} Townsley, L., Broos, 
P., Bauer, F., 
\& Getman, K.\ 2003, Bulletin of the American Astronomical Society, 35, 644 

\bibitem[Townsley et al.(2011a)]{CCCP-intro} Townsley, L.~K., 
Broos, P.~S., Corcoran, M.~F., et al.\ 2011, \apjs, 194, 1 

\bibitem[Townsley et al.(2011b)]{CCCP-diffuse1} Townsley, L.~K., 
Broos, P.~S., Chu, Y.-H., et al.\ 2011, \apjs, 194, 15 

\bibitem[Townsley et al.(2011c)]{CCCP-diffuse2} Townsley, L.~K., 
Broos, P.~S., Chu, Y.-H., et al.\ 2011, \apjs, 194, 16 

\bibitem[Walborn(1982)]{Walborn82} Walborn, N.~R.\ 1982, \aj, 87, 1300 

\bibitem[Weisskopf et al.(2002)]{Weisskopf02} Weisskopf, M.~C., 
Brinkman, B., Canizares, C., et al.\ 2002, \pasp, 114, 1 

\bibitem[Wang et al.(2011)]{CCCP-Tr15} Wang, J., Feigelson, E.~D., Townsley, L.~K., et al.\ 2011, \apjs, 194, 11 

\bibitem[Wolk et al.(2005)]{Wolk05} Wolk, S.~J., Harnden, 
F.~R., Jr., Flaccomio, E., et al.\ 2005, \apjs, 160, 423 

\bibitem[Wolk et al.(2011)]{CCCP-Tr16} Wolk, S.~J., Broos, P.~S., Getman, K.~V., et al.\ 2011, \apjs, 194, 12 

\end{thebibliography}
\end{document}